\documentclass{jfm}
\usepackage{graphicx}
\usepackage{epstopdf, epsfig}
\usepackage{hyperref}
\usepackage{subcaption}
\usepackage{amsmath}
\usepackage{comment}
\usepackage[linesnumbered, ruled, lined]{algorithm2e}

\shorttitle{Generating wall-bounded turbulent inflows at high Reynolds numbers}
\shortauthor{R. Stanly, T. Mukha, M. Karp, S. Markidis and P. Schlatter}
\title{Generating wall-bounded turbulent inflows at high Reynolds numbers}
\author{Ronith Stanly\aff{1}
  \corresp{\email{ronith@kth.se}},
  Timofey Mukha\aff{2},  
  Martin Karp\aff{1},
  Stefano Markidis\aff{3},
 \and Philipp Schlatter\aff{1,4}
 }

\affiliation{\aff{1}Department of Engineering Mechanics, KTH Royal Institute of Technology, SE-100 44 Stockholm, Sweden.
\aff{2}Computer, Electrical and Mathematical Sciences and Engineering Division, King Abdullah University of Science and Technology (KAUST), 23955-6900 Thuwal,  Kingdom of Saudi Arabia.
\aff{3} Division of Computational Science and Technology, EECS, KTH Royal Institute of Technology, SE-100 44 Stockholm, Sweden.
\aff{4} Institute of Fluid Mechanics (LSTM), Friedrich--Alexander--Universit\"at Erlangen--N\"urnberg (FAU), DE-910 58 Erlangen, Germany.
}

\begin{document}

\maketitle

\begin{abstract}

One of the main challenges in simulating high Reynolds number ($\Rey$) turbulent boundary layers (TBLs) is the long streamwise distance required for large-scale outer-layer structures to develop, making such simulations prohibitively expensive. We propose an inflow generation method for high $\Rey$ wall turbulence that leverages the known structure and scaling laws of TBLs, enabling shorter development lengths by providing rich input information. As observed from the inner-scaled pre-multiplied spectra of streamwise velocity, with an increase in $\Rey$ the outer region grows and occupies more of the spanwise wavenumber space in proportion to the increase in $\Rey$; while the inner region remains approximately the same. Exploiting this behavior, we generate high-$\Rey$ inflow conditions for a \textit{target} $\Rey$ by starting from cross-stream velocity slices at a lower \textit{base} $\Rey$. In spectral space, we identify the inner and outer region wavenumbers, and shift the outer-region components proportionally to the desired $\Rey$ increase. We closely examine the capability of this method by scaling a set of velocity slices at $Re_\theta=2240$ and 4430 to $Re_\theta=8000$, and using them as inflow conditions for direct numerical simulations (DNS) of spatially developing TBLs growing from $\Rey_\theta=8000-9000$. The skin friction coefficient and shape factor predicted by the new method, regardless of the \textit{base $\Rey$} tested, is within $\pm3.5\%$ and $\pm0.5\%$, respectively, of that of a precursor simulation right from the inlet. Reynolds stresses match very well after approximately $8~\delta_{99_0}$. This gives an order of magnitude reduction in development length compared to other methods proposed in the literature.

\end{abstract}

\begin{keywords}
Authors should not enter keywords on the manuscript, as these must be chosen by the author during the online submission process and will then be added during the typesetting process (see http://journals.cambridge.org/data/\linebreak[3]relatedlink/jfm-\linebreak[3]keywords.pdf for the full list)
\end{keywords}

\section{Introduction}

Improving our fundamental understanding of high Reynolds number ($\Rey$) turbulent boundary layers require advancements in both experimental methods as well as simulation techniques. This would, for instance, enable us to answer questions such as the extent of applicability of Taylor's hypothesis by knowing the scale-dependent distribution of convection velocities or to unravel the influence of large scale structures on near-wall turbulence or to find out the differences between canonical cases of wall turbulence, viz., pipes~\citep{Massaror_pipe-2024}, channels~\citep{MONTY_pipe_channel-2007,MONTY-pipe_channel_TBL-2009} and turbulent boundary layers~\citep{HUTCHINS_MARUSIC_2007,high_re_tbl_review-smits-2011}. While improving the spatial resolution of sensors is one of the known challenges in advancing experimental methods for wall turbulence, we keep the focus of this current work on simulation methods for spatially developing turbulent boundary layers, where the turbulence inflow condition is often a limiting factor. Although alternative methods that simulate a temporally developing boundary layer instead~\citep{spalart_1988,temporal_TBL-CnF-Biau2023,temporal_TBL-pasha-2024} circumvent this issue by enabling streamwise periodicity, this current work is aimed at advancing inflow conditions for spatially developing boundary layers. A good turbulent inflow method is expected to develop the flow to the required turbulent state in as short streamwise distance as possible~\citep{Tabor_2010}. This becomes difficult to achieve as one goes to higher $\Rey$ as the larger structures in the outer region take longer in, time and space, to develop~\citep{sillero_2013}. 
\citet{wu_2017} and~\citet{lyrintzis_inflow_review-2018} provide a concise  overview of several methods that can be used to provide an inflow condition. In principle, we can distinguish two basic types:  i) methods that re-use real turbulent fields after some transformation and ii) methods that create artificial turbulence, and mixtures of the two.   
Some of the more commonly used  methods specifically for turbulent boundary layers include precursor-based methods~\citep{malm_tubr_inflow,malm_tubr_inflow_journal}, recycling-rescaling methods~\citep{Lund_1998} and volume force-based (tripping) methods~\citep{schlatter_örlü_2012}. \\%The development lengths of some of these methods were recently studied~\citep{Stanly_Du-JFM-2024}.\\ 

Among the different categories of inflow methods, some synthetic methods utilize the known structure of the coherent eddies~\citep{coherent_structure-Robinson-1991} within a turbulent boundary layer. Several of these follow directly or indirectly from the works of~\citet{Perry_Chong_1982},~\citet{Perry_Marusic_1995} and~\citet{Marusic_Perry_1995}, who demonstrated using the attached eddy hypothesis of \citet{townsend1956,townsend1976} that the first and second-order statistics of a turbulent boundary layer can be reproduced using a superposition of these coherent structures. A good review of attached eddy model is given by~\citet{attached_eddy_review-marusic-2019} and further enhancements to this model were later pursued~\citep{dileep_chandran_2020,Deshpande2021}.~\citet{attached_eddy_turb_inflow-Marusic2006} applied this idea to create an inflow condition where they assumed the flow was populated by hairpin vortices with a given population density. Biot-Savart law was used to compute the velocity field these vortices induced around them  and image vortices at the wall were used to enforce no penetration condition. However, they mentioned that it ``would be impractical to resolve" the inner scale structures ``at moderate to high Reynolds number" flows using this idea ``on small/medium sized grids" and that the method was better suited for LES or hybrid RANS-LES which did not resolve the wall. ~\citet{Inflow_inner_outer-Sandham-2003} proposed a simple to implement, coherent structure based model which split the boundary layer into two regions: inner region populated with lifted streaks and an outer region populated with three dimensional eddies. Their deterministic model introduced disturbances corresponding to the inner and outer regions along with their associated phase information.~\citet{touber2009} later evaluated this method and found that it generated a mean velocity profile with an unrealistic outer region, as well as resulted in a spurious secondary peak in turbulence intensity and Reynolds shear-stress profiles in the outer region. These were possible to be corrected by running several precursors and tuning certain parameters like~\citet{Sndberg_Sandham_2008} who applied a modified version of this method to simulate the sensitive case of noise from a boundary layer passing over a trailing edge and~\citet{Sandham_Sandberg-2009} who used it to simulate a turbulent mixing layer. In another work, \citet{synthetic_eddy_wall_info-pierre_sagaut-2009} proposed a modification to the synthetic eddy method by~\citet{Jarrin_2006} for large eddy simulation of spatially developing boundary layers. They split the random structures into structures belonging to different modes and included their proper length, time, velocity and vorticity contents, to better represent the vortex structures inside a turbulent boundary layer in the wall-normal direction. With this they were able to better specify the information of the modes for the buffer and logarithmic layers which helped reduce their development length.\\

\begin{figure}
    \centering
    \includegraphics[width=0.5\linewidth]{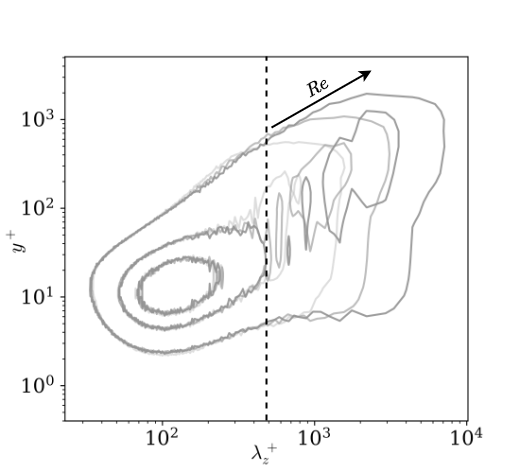}
    \caption{Pre-multiplied spectra of streamwise velocity $u$ at increasing $\Rey$ ($\Rey_\theta=2240, 4430, 8000$; contour lines of increasing darkness show increasing $\Rey$), \emph{i.e.,} $E_{uu}(k_z)\cdot k_z /u_\tau^2$. $^{+}$ indicates normalization using inner units of the respective $\Rey$ and $y$ is the wall-normal coordinate. Notice that the inner region (left of the vertical dashed line) remains the same, whereas the outer region (right of the vertical dashed line) grows with increasing $\Rey$. For each level of darkness, contour lines are plotted for levels $[0.5, 1.5,  2.5, 4.0]$. Data from \cite{EitelAmor_2014}.}
    \label{fig:graphical_abs}
\end{figure} 

\begin{figure}
    \centering
    \begin{subfigure}[b]{6.5cm}
        \centering                
        \includegraphics[width=6.5cm]{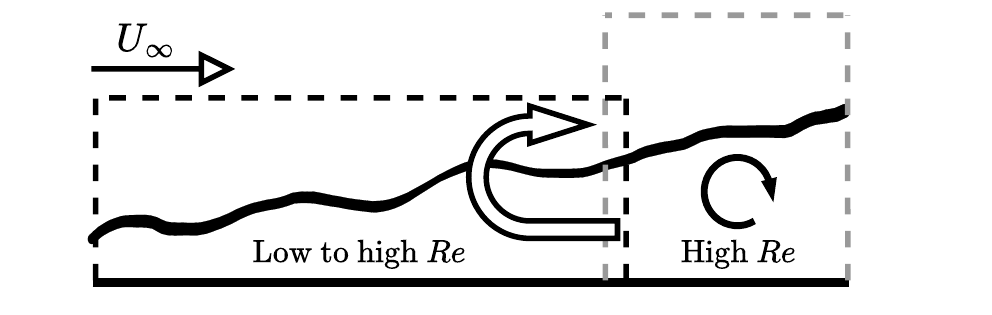}
        \caption{Classic precursor inflow}
        \label{fig:schm_precursor}
    \end{subfigure}
    \begin{subfigure}[b]{6.5cm}
        \centering
        \includegraphics[width=6.5cm]{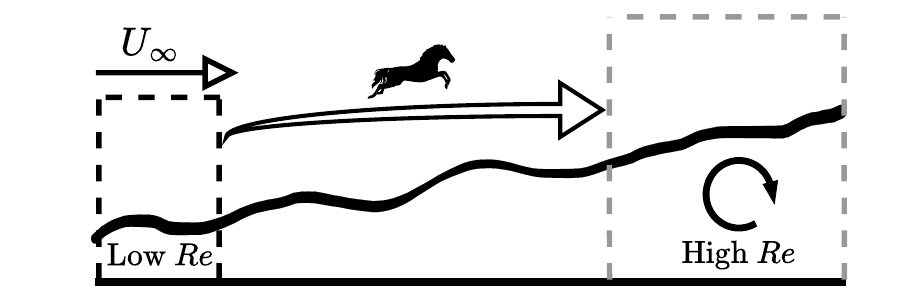}
        \caption{Proposed method}
        \label{fig:schm_scaling}
    \end{subfigure}
    \caption{Schematic of the proposed method highlighting its potential to reduce computational cost by up-scaling low-$\Rey$ precursor data to high $\Rey$.
(a) In classic precursor or existing methods, the precursor domain must grow with the \textit{target} $\Rey$.
(b) The proposed scaling method enables a small, fixed-size precursor domain regardless of \textit{target} $\Rey$.
Black dashed boxes show precursor domains; grey dashed boxes show main TBL simulation domains.
Single-lined arrows indicate flow direction, double-lined arrows indicate Dirichlet inflow application. The jumping horse symbolizes the proposed scaling-leap bypassing the costly development of large-scale structures.} %The jumping horse symbolizes the proposed scaling procedure, which enables bypassing the computationally expensive development of large-scale structures.}
    \label{fig:sch_prec_vs_scaling}
\end{figure}

In this work, we propose a structure-based inflow method whose core idea can be summarized using Figure~\ref{fig:graphical_abs}. %depicts the core idea of the method. 
As observed from the inner-scaled spectra of streamwise velocity pre-multiplied with spanwise wavenumbers (interested readers are referred to the work of \citet{jimenez1998largest} for a discussion on such spectra), with an increase in $\Rey$ the outer region grows and occupies more spanwise wavenumbers as opposed to the inner region which almost remains the same. Using the proposed method, time-dependent cross-stream velocity slices of a turbulent boundary layer at a given $\Rey$ are first split into inner and outer regions in spectral space, followed by scaling (in space, time and energy) the spanwise wavenumbers corresponding to %the inner region in inner units and those corresponding to 
the outer region appropriately %in outer units 
%then the inner region is scaled in inner units and the outer region in outer units 
to then reconstruct velocity fields at any higher $\Rey$ of interest.
%and then appropriately scaled to attain any higher $\Rey$ of interest.
%which uses the growth rate obtained from attached eddy hypothesis to scale the cross-stream velocity slices of a turbulent boundary layer at a give Re to obtain the same at any higher Re. 
As schematically shown in Figure~\ref{fig:sch_prec_vs_scaling}, such a procedure enables a leap from low $\Rey$ precursor data to much higher $\Rey$. This avoids the need for prohibitively expensive precursor domains, which must be larger the higher the target $\Rey$ one aims to simulate. 

In the remainder of this paper, we first describe the  scaling method in detail in \S\ref{sec:methodology}, followed by demonstrating its application as inflow condition in \S\ref{sec:results}. Final conclusions and outlook are given in \S\ref{sec:discuss}.

\section{Methodology}
\label{sec:methodology}

To obtain velocity fields at a target high $\Rey$ (we call this the ``\textit{target $\Rey$}"), the proposed method requires the following two data-sets as inputs:

\begin{itemize}
    \item time-dependent cross-stream velocity slices at a lower $\Rey$ (we call this the ``\textit{base $\Rey$}")
    \item numerical value of the $\textit{target}~\Rey$. %mean velocity profiles, $U_{i,~\text{target}}(y)$, at the \textit{target $\Rey$}.
\end{itemize}

And, if not a generic zero-pressure-gradient turbulent boundary layer, appropriate scaling behaviour of the velocity spectral density and velocity profiles.

Using the $\textit{target}~\Rey$, the following derived inputs can be calculated: mean velocity profiles $U_{i,~\text{target}}(y)$ at the \textit{target $\Rey$}, the scaling factor for space and time $sc$, energy scaling factor $E_{sc}$. The subscript $i$ denotes the three components of velocity (streamwise, wall-normal and spanwise, respectively), although the spanwise component is identically zero due to the spanwise homogeneity of turbulent boundary layers.

The velocity fields at the \textit{base $\Rey$} can be obtained by running a low cost precursor, or simply be taken from an existing database. These are ideally wide enough in the spanwise direction to incorporate the wider structures that would appear after applying the new scaling procedure. Potential alternatives to avoid a very wide domain for the precursor are discussed in \S\ref{sec:discuss}. This \textit{base $\Rey$} should have some large-scale structures present beyond an inner-scaled spanwise wavelength of $\lambda_z^+>500$, as seen on a pre-multiplied spectra of streamwise velocity (see Figure~\ref{fig:graphical_abs}). In other words, the \textit{base $\Rey$}  should be high enough for outer scaling to be valid.
% which is preferably near or above the $\Rey$ regime
%sufficient scale separation. Although~\citet{MARUSIC2010} mentions an $Re_{\tau}>1700$ and~\citet{alfredsson_outer_peak-2011} suggests an $Re_\tau>2000$ as necessary to see sufficient scale separation (as evidenced by an emerging outer peak in the pre-multiplied spectra of streamwise velocity), in \S\ref{sec:results} we demonstrate that the current method works well even for much lower $\Rey$ ($Re_{\theta}=2240$, \emph{i.e.}, $Re_{\tau}\approx 740$). 
 In this work, we use the database of~\citet{EitelAmor_2014}, where finely-resolved large eddy simulation was performed upto a Reynolds number based on momentum-loss thickness of $Re_\theta=8300$ and time series data of cross-stream velocity slices sampled on a coarser $y$-grid (of 47 points) is available at $Re_\theta=2240, 3740, 4430, 5740$, and $8000$. As what will be shown in subsequent sections, using a \textit{base \Rey$_\theta=2240$} was sufficient for the proposed method to work seamlessly. On the other hand, an extreme case of using a very low \textit{base}  $\Rey_\theta=790$ needed hand-tuning of the scaling parameters for space and energy by examining the pre-multiplied spectra of streamwise velocity. This is because outer scaling does not hold true for such a low \Rey ~where viscous effects and interactions between the inner and outer regions are strong.
 
 The mean streamwise velocity profile at any \textit{target $\Rey$} can be obtained by using, for instance, the log-law~\citep{vonKarman1930}, Cole's law~\citep{coles1956}, composite profile~\citep{monkewitz2007} or any other suitable universal profile~\citep{universal_profile-Subrahmanyam_Cantwell_Alonso_2022}. On the other hand, the mean wall-normal velocity profile can be obtained from methods like the ones proposed by~\citet{scaling_mean_V_ZPG-2023}  or by using the continuity equation along with self-similarity to derive approximations, for instance. For the numerical experiments performed in this work, since the database contained the information at the \textit{target $\Rey$}, we use the mean velocity profiles as well from the database. 

In terms of the key operations involved in this method, performing Fourier transform allows us to extract the inner and outer structures, to then scale in space. Subsequently, performing Singular Value Decomposition (SVD) in the vertical direction $y$ allows us to get access to the dominant modes in $y$ along with their time evolution, via the time coefficients. These time coefficients can then be used to scale the temporal evolution of those selected structures accordingly. Performing SVD enables two extra possibilities: (i) to create a low-order model for the wall-normal dynamics, and (ii) to incorporate synthetically extended time coefficients generated by vector autoregression as described by~\citet{Stanly_Du-JFM-2024}, which allows the time signal of turbulent fields scaled to high $\Rey$ by the proposed method to also be extended to arbitrary lengths in time. Otherwise the scaling procedures mentioned in this work can also be performed just after a Fourier transformation of time-dependent cross-stream velocity fields. Details of the scaling procedure are explained in the following subsections.

\subsection{Extracting inner and outer structures}
\label{sec:method_extract}

In order to extract structures corresponding to the inner and outer regions from the velocity slices at the \textit{base $\Rey$}, we take advantage of the uniform grid along the spanwise ($z$) periodic direction and perform Fourier transform, by way of Discrete Fourier Transform (DFT), on the three components of velocities along this direction. This can we written in a continuous sense as

\begin{equation}\label{eq:z-fourier}
%\hat{u}_i^{(k_z^{\text{phys}})}(y,t) = \frac{1}{L_z} \int_0^{L_z} u_i(y,z,t)\, e^{-i k_z^{\text{phys}} z} \, dz
\hat{u}_i(k_z^{\text{phys}}, y, t) = \frac{1}{\sqrt{L_z}} \int_0^{L_z} u_i(y,z,t) e^{-i k_z^{\text{phys}} z} dz
\end{equation}

%\begin{equation}\label{eq:z-fourier}
%    u_i(y,z,t) = \sum\limits_{k_z=-n_z/2+1}^{n_z/2} \hat{u_i}^{(k_z)}(y,t)e^{2\pi i k_zz/L_z}\,,
%\end{equation}

where the subscript $i$ varies from $1$ to $3$ representing three components of the velocity vector, $k_z^{\text{phys}}$ are the physical wavenumbers in $z$, $L_z$ is the domain length in the $z$ direction and ${\hat{}}$ denotes the corresponding variable in Fourier space obtained through Fourier transform. The discrete, unitary version of the same can be written as

\begin{equation}
\label{eq:z_dft}
%\hat{u}_i^{(k_z)}(y,t) = \frac{1}{n_z} \sum_{j=0}^{n_z-1} u_i(y,z_j,t)\, e^{-2\pi i k_z j / n_z}
\hat{u}_i^{(k_z)}(y,t) = \frac{1}{\sqrt{n_z}} \sum_{j=0}^{n_z-1} u_i(y,z_j,t) e^{-2\pi i k_z j / n_z}.
\end{equation}

This normalization using $\sqrt{n_z}$ ensures that the Parseval relation holds~\citep{fft_orthonormal_bounchaleun2019,trefethen2000spectral}.
%This normalization using $\sqrt{n_z}$ ensures each Fourier mode has unit energy,
%%unitary energy preservation,
%i.e., the Parseval relation holds~\citep{fft_orthonormal_bounchaleun2019,trefethen2000spectral}. 
%In other words, this makes the Fourier modes ortho-normal. Such energy consistency and orthonormality preservation across domains (e.g., between physical and spectral spaces) is important as we work with POD and spectral reconstructions where energy and inner products are key. 
If only one half of the complex conjugates are later retained, like what we do, the factor should be $\sqrt{n_z/2}$ instead of $\sqrt{n_z}$. Here the wavenumber index $k_z$ is related to the physical wavenumber $k_z^{\text{phys}}$ as
\begin{equation}
k_z^{\text{phys}} = \frac{2\pi k_z}{L_z}.
\end{equation}

In practice, we take $k_z$ in the range from $0$ to $n_z/2$, assuming $n_z$ is even. This makes the normalization factor in Equation~(\ref{eq:z_dft}) as $\sqrt{n_z/2}$. Because the velocity field is real-valued, we exploit the Hermitian symmetry of the DFT and retain only one half of the wavenumbers
\begin{equation}
k_z \in \left[ 0, 1, 2, \dotsc, \frac{n_z}{2} \right].
\end{equation}

Following this, spanwise wavenumbers corresponding to the inner and outer regions of the turbulent boundary layer are extracted depending on their corresponding value of ${\lambda_z}^+$ which calculated as
\begin{equation}
    {\lambda_z}^+ = \frac{2\pi}{k_{z}l^{*}},
\end{equation}
where $l^*$ is the viscous length-scale obtained as $l^*=\nu/u_{\tau}$. Here, $u_{\tau}$ is the friction velocity calculated using wall-shear stress ${\tau}_w$ and density $\rho$, as $u_{\tau}=\sqrt{{\tau}_w/\rho}$.
Following this, as observed from Figure~\ref{fig:graphical_abs}, a range of $k_z$ values corresponding approximately to $30 \leq{\lambda_{z}}^+\leq 500$ are extracted to obtain the inner region (\emph{i.e.,} $\mathcal{K}_{\text{inner}}$). For extracting the outer region, relevant range of $k_z$ (\emph{i.e.,} $\mathcal{K}_{\text{outer}}$), as observed from the corresponding pre-multiplied spectrum, are extracted depending on the $\Rey$ of the available dataset. For instance for $\Rey_{\theta}=2240$, $k_z$ corresponding approximately to $500 <{\lambda_{z}}^+\leq 1700$; and for $\Rey_{\theta}=4430$, $k_z$ corresponding approximately to $750 <{\lambda_{z}}^+\leq3000$ are extracted for the outer region.

These realizations $\hat{u_i}^{(k_z)}(y,t)$ of the selected wavenumbers ($k_z$) are stacked into adjacent columns to construct the complex snapshot matrix. %for each selected wavenumber, $k_z$. 
SVD is then performed in the wall-normal direction ($y$) for each of those wavenumbers. This results in time coefficients, $a^{(k_{z},n)}(t)$ of the $n$-th proper orthogonal decomposition (POD) mode, and complex POD modes, $\hat{\varphi_i}^{(k_z,n)}(y)$. The above operations result in an expansion as follows 
\begin{equation}\label{eq:POD_u}
    \hat{u_i}^{(k_z)}(y,t) \approx \sum\limits^{n_{\rm modes}-1}_{n=0}a^{(k_z,n)}(t)\hat{\varphi_i}^{(k_z,n)}(y).
\end{equation}

\subsection{Scaling procedure }
\label{sec:method_scaling}
Overall, scaling the velocity fields in $\Rey$ involves shifting the spanwise wavenumbers ($k_z$) of the selected structures (to scale the structures in $z$) and scaling the relevant modes in $y$, $t$ as well as scaling their energy. In principle, the inner region is to be scaled in inner units and the outer region in outer units. However, %as can be observed from the growth of skin friction ($c_f$) with $\Rey_{\theta}$ in Figure ----, 
it is known, for instance from~\citet{EitelAmor_2014}, that the growth of the structures in inner region of a turbulent boundary layer is almost negligible compared to the growth of the outer region as we go to higher $\Rey$. Moreover, the fastly evolving near-wall structures would compensate for these small differences quite rapidly, as opposed to the slow outer structures. For instance, when comparing $\Rey_\theta=2240$ to $\Rey_\theta=8000$, the viscous length scale $l^*$ increases by a factor of $1.16$, whereas the boundary layer thickness $\delta_{99}$ increases by a factor of $3.96$. Owing to this, in this work, the inner region from the \textit{base $\Rey$} is kept as such, and only the outer region is scaled in outer units in our scaling procedure to reach the \textit{target $\Rey$}.

For the selected range of $k_z$, corresponding to the outer region, scaling is performed using the scaling factor, $sc$, found out as
\begin{equation}
    sc = \frac{{\delta_{99}}_{\text{target}}}{{\delta_{99}}_{\text{base}}},
\end{equation}
where ${\delta_{99}}_{base}$ is the boundary layer thickness of the velocity slices available from the database and ${\delta_{99}}_{target}$ is the boundary layer thickness at the \textit{target $\Rey$} calculated using the mean velocity profile obtained as input. This scaling factor is 1.9 for the case where velocity slices at $Re_\theta=4430$ are up-scaled to $Re_\theta=8000$, and 3.9 when $Re_\theta=2240$ are scaled up to $Re_\theta=8000$. 

\subsubsection{Scaling in $z$}
\label{sec:shifting_in_kz}
Scaling in the spanwise direction $z$ is performed by means of shifting each of the $n$ required POD modes (depending on how much percentage of the turbulent kinetic energy (TKE) is intended to be retained in the reduced representation) of the selected spanwise wavenumbers corresponding to the outer region, $\mathcal{K}_{\text{outer}}$, to the new position in spectral space. In this work, the first 40 POD modes for each of the selected wavenumbers were kept. This retained about $97\%$ of TKE across the retained modes. Based on our previous work~\citep{Stanly_Du-JFM-2024}, retaining this much of TKE in the number of POD modes had no noticeable negative effects due to truncation. Scaling in $z$ is performed as follows and is illustrated with an example in Figure~\ref{fig:scaling_z}
\begin{equation}
 %a^{(\text{round}(k_z/sc),n)} = a^{(k_z,n)}  
 \hat{\varphi_i}^{(\text{round}(k_z/sc),n)}(y) = \hat{\varphi_i}^{(k_z,n)}(y).
\end{equation}

From here, we use $\mathcal{K}_{\text{outer}}$ to refer to the newly scaled/shifted range of outer $k_z$, $\emph{i.e., }$ $\mathcal{K}_{\text{outer}}\approx \text{round}(\mathcal{K}_{\text{outer}}/sc)$.

%\vspace{2ex}
%\textit{POD modes to physical space:}
%\vspace{1ex}

\subsubsection{Reconstruction of POD modes to physical space}
%At this stage, before performing further scaling operations, the POD modes can be brought to physical space by performing an independent inverse Fourier transform on each POD mode of each $k_z\in (\mathcal{K_{\text{inner}}} \cup~\mathcal{K_{\text{outer}}})$ as:

%\begin{equation}\label{eq:z-inv_fourier}
%    \varphi_i^{(kz,n)}(y,z) = \sum_{k_z~\in~(\mathcal{K}_{\text{inner}}~\cup~\mathcal{K}_{\text{outer}})} \hat{\varphi_i}^{(k_z,n)}(y)e^{2\pi i k_zz/L_z}
%\end{equation}

Before applying further scaling operations, each POD mode can be independently reconstructed in physical space by applying an inverse Fourier transform for each wavenumber $k_z \in (\mathcal{K}_{\text{inner}} \cup \mathcal{K}_{\text{outer}})$. This can be shown in a continuous form as
\begin{equation}\label{eq:z-inv_fourier}
    \varphi_i^{(k_z,n)}(y,z) = \frac{2}{\sqrt{n_z/2}} \hat{\varphi}_i^{(k_z,n)}(y)\, e^{i k_z^{\text{phys}} z},
\end{equation}
and equivalently in the discrete form as
\begin{equation} \label{eq:idft_phi}
\varphi_i^{(k_z,n)}(y, z_j) = \frac{2}{\sqrt{n_z/2}} \hat{\varphi}_i^{(k_z,n)}(y) e^{2\pi i k_z j / n_z},
\end{equation}
where the factor $2$, within and outside the square-root, comes in as we retain only one half of the wavenumbers that exhibit Hermitian symmetry. The factor $\frac{1}{\sqrt{n_z/2}}$ is the unitray normalization to undo the normalization performed in Equation~(\ref{eq:z_dft}). The factor $2$ outside the square-root can be removed if $k_z \in (0,\frac{n_z}{2})$ as they have no complex conjugate pairs.

\subsubsection{Scaling in $y$}

Following this, each of the selected POD modes are to be scaled in the wall-normal direction $y$, as the new structures would grow in $y$ as well, when appearing at the higher $\Rey$. %For this, the spatial extent of these modes are upscaled in $y$ by multiplying with the same scaling factor ($sc$). 
To achieve this, while keeping the POD modes the same, the y axis is upscaled such that
\begin{equation}
    y_{\text{target}} = y \cdot sc.
\end{equation}

If one would now plot the POD modes using this new $y_{\text{target}}$ axis, the modes will look like it has grown in the wall-normal coordinate. However, we need these grown modes on the same initial grid ($\emph{i.e.,}~y$). To get this, the next step is to chop-off the $y_{\text{target}}$ values beyond $\text{max}(y)$ such that both meshes go up to the same height. This would result in $\text{size}(y_{\text{target}})<\text{size}(y)$, in other words $y_{\text{target}}$ ends up being coarse than $y$. To get back the POD modes to the initial $y$ grid from $y_{\text{target}}$, a 2-D spline interpolation is now performed

\begin{equation}
    {\varphi_i}^{(k_z,n)}(y_{\text{target}},z) \rightarrow {\varphi_i}^{(k_z,n)}(y,z).
\end{equation}

\begin{figure}
    \centering
    \begin{subfigure}[b]{6.5cm}
        \centering                
        \includegraphics[width=6.5cm]{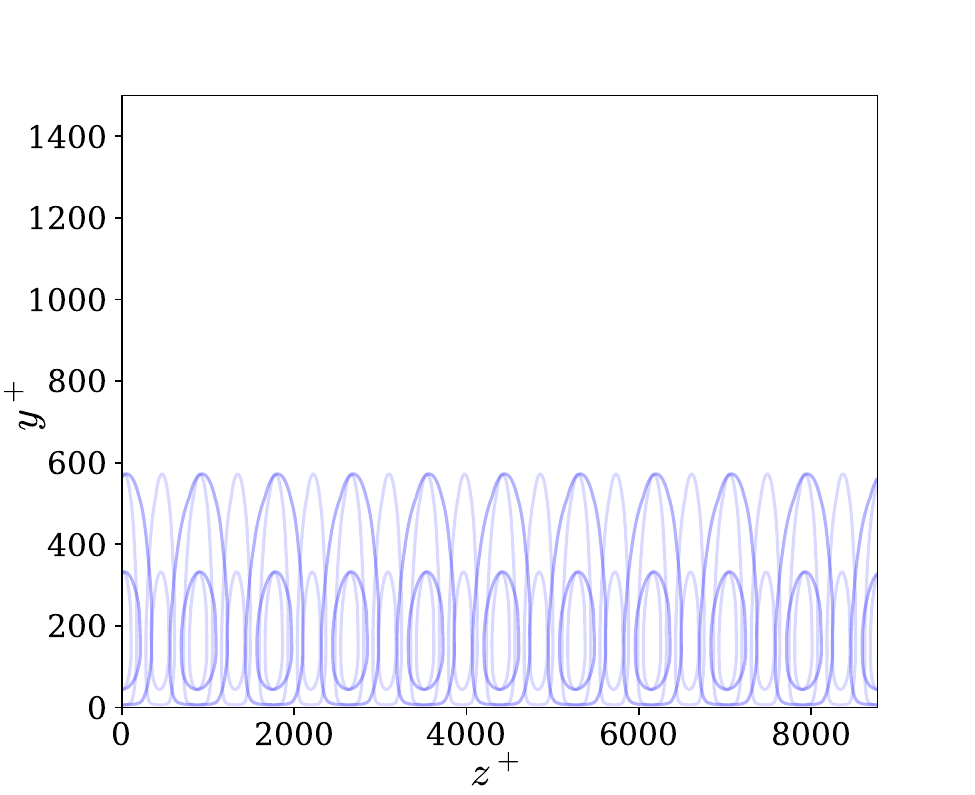}
        \caption{Scaling in $z$}
        \label{fig:scaling_z}
    \end{subfigure}
    \begin{subfigure}[b]{6.5cm}
        \centering
        \includegraphics[width=6.5cm]{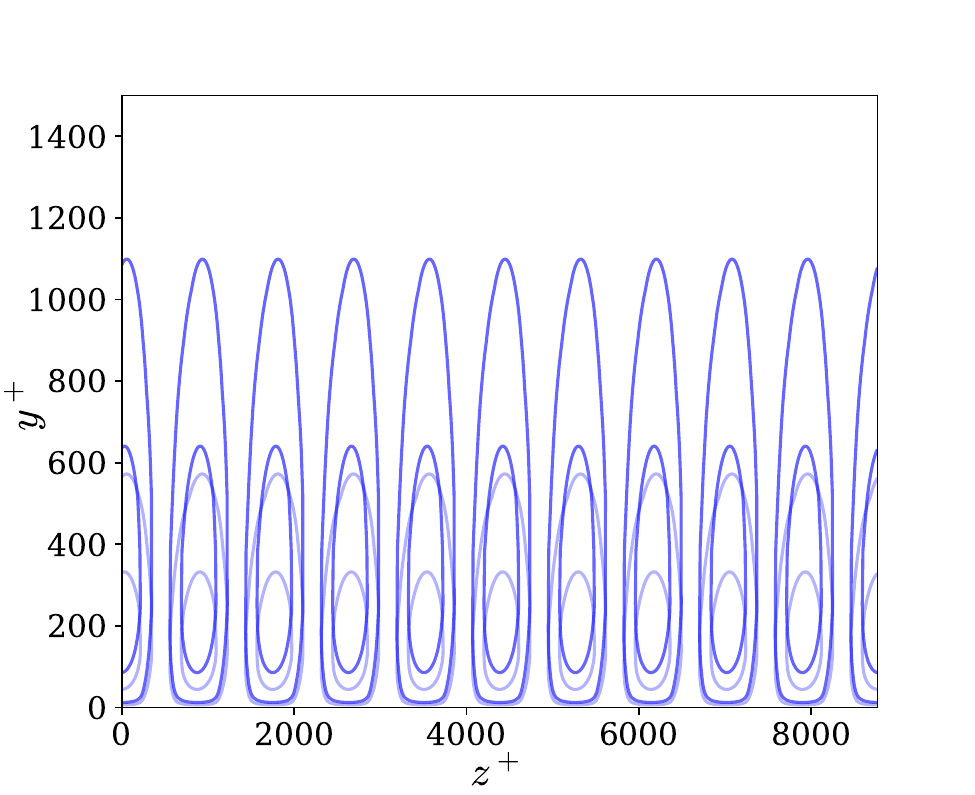}
        \caption{Scaling in $y$}
        \label{fig:scaling_y}
    \end{subfigure}
    \caption{Effect of scaling in $k_z$ and $y$ shown using the real part of the 0th POD mode of $k_z=10$ of streamwise component of velocity  of \textit{re4k\_sc}, $\emph{i.e.,}~\text{Real}\{\varphi_1^{(k_z=10,n=0)}\}$. In both (a) and (b), the lighter shade of blue shows the state before scaling, and the darker shade shows the state after scaling. For both, contour lines are plotted for levels $[\pm0.0015, \pm0.0030, \pm0.0045]$.} 
    \label{fig:scaling_z_y}
\end{figure}

This scaling in $y$ can be seen in Figure~\ref{fig:scaling_y}.

\subsubsection{Scaling in $t$}

Next, the time coefficients have to be adjusted as the newly produced bigger structures in the outer region are expected to convect slower than the initial smaller counterparts. For this, the time array is downscaled by $sc$ and then the new time coefficients corresponding to this new time array is interpolated from the old time coefficients. This can be shown as
\begin{equation}
    t_{\text{target}} = t / sc,
\end{equation}
and for the coefficients
\begin{equation}
    a^{(k_z,n)}(t) \rightarrow a^{(k_z,n)}(t_{\text{target}}).
\end{equation}

\subsubsection{Scaling the energy}
\label{sec:method_scaling_energy}
For upscaling the energy of the outer structures, the square of the scaling factor (${E_{sc}}^2$) is found out between the mean streamwise turbulence intensity ($u'u'$) from the log and outer regions (say, $y^+>100$) of the \textit{base $\Rey$} obtained from the database, and the mean streamwise turbulence intensity at the \textit{target $\Rey$} in the same region (\emph{i.e.,} $y^+>100$). For the latter, the following expression proposed by~\citet{alfredsson_outer_peak-2011} is used
\begin{equation}
    \frac{u_{rms}(y)}{U(y)} = 0.031 + 0.260 \left(  1-\frac{U(y)}{U_{\infty}} \right),
\end{equation}
where $U(y)$ is the mean streamwise velocity profile at the \textit{target $\Rey$} and $U_{\infty}$ is the mean freestream velocity. Although this scaling factor for energy corresponds to the streamwise fluctuations, due to lack of other estimates to find scaling factors for fluctuations in the wall-normal and spanwise directions (and following some tests) the same factor is applied to upscale the energy of the wall-normal and spanwise fluctuations as well. This is done by multiplying this energy scaling factor to the corresponding outer modes of the three velocity components in the reconstruction phase. This scaling factor for energy is found to be $E_{sc}\approx1.3$ for the case where velocity slices at $Re_\theta=4430$ are up-scaled to $Re_\theta=8000$, and $E_{sc}\approx1.5$ when $Re_\theta=2240$ is scaled up to $Re_\theta=8000$. 

\subsection{Reconstruction}
After applying the scaling procedure, the mean velocity profiles at the \textit{target $\Rey$}, $U_{i,~\text{target}}(y)$, (obtained as mentioned in the beginning of \S\ref{sec:methodology}) are used along with the selected inner modes and the scaled outer modes to reconstruct the velocity fields, $u_{i,~\text{target}}$, at the \textit{target $\Rey$}. Note that the scaling of energy of the outer modes using the scaling factor $E_{sc}$ is performed at this stage. The reconstruction of the time dependent two-dimensional velocity snapshots in the $yz$-plane can be shown as
%
\begin{comment}
\begin{equation}\label{eq:POD_rec_orig}
    u_{i,~\text{target}}(y,z,t) = \sum\limits_{k_z=-n_z/2+1}^{n_z/2} \sum\limits^{n_{\rm modes}-1}_{n=0}a^{(k_z,n)}(t)\hat{\varphi_i}^{(k_z,n)}(y)e^{2\pi i k_zz/L_z}.
\end{equation}
\end{comment}

\begin{equation}\label{eq:POD_rec_orig}
\begin{aligned}
%&\text{Reconstruct velocity fields at \textit{target} } \Rey:\\
u_{i,~\text{target}}(y,z,t_{\text{target}}) =
&\quad U_{i,~\text{target}}(y,z) \quad \text{(for } k_z = 0,~n = 0\text{)} \\
&\quad + \sum_{k_z\in \mathcal{K}_{\text{inner}}} \sum_{n=0}^{n_{\text{modes}} - 1} a^{(k_z,n)}(t) \varphi_i^{(k_z,n)}(y,z)\\
&\quad + \sum_{k_z\in \mathcal{K}_{\text{outer}}} \sum_{n=0}^{n_{\text{modes}} - 1} E_{sc} \cdot a^{(k_z,n)}(t_{\text{target}}) \varphi_i^{(k_z,n)}(y,z).
\end{aligned}
\end{equation}

This entire process is outlined in Algorithm~\ref{algo_whole}, where $\mathcal{F}_z$ and $\mathcal{F}_z^{-1}$ represent implementations of the fast Fourier transforms, such as the Matlab operations \texttt{fft()} and \texttt{ifft()}, \emph{i.e.}, forward and inverse transform, respectively. The normalizations performed on them (\emph{i.e.,} on the output from these functions), in steps 1 and 11 using the factor $\sqrt{\frac{n_z}{2}}$, yield Equation~(\ref{eq:z_dft}) and (\ref{eq:idft_phi}) after accounting for the normalization within these Matlab functions.

\SetAlgoBlockMarkers{begin}{end}  % replaces {} block with explicit begin...end
\SetKwBlock{AllLoop}{\textbf{for each} $k_z \in [0, {n_{z}}-1]$ \textbf{do}}{\textbf{end do}}
\SetKwBlock{OuterLoop}{\textbf{for each} $k_z \in \mathcal{K}_{\text{outer}}$ \textbf{do}}{\textbf{end do}}
\SetKwBlock{funcscalez}{\textbf{Scale in $z$:}}{} %\textnormal
\SetKwBlock{funcscaley}{\textbf{Scale in $y$:}}{}
\SetKwBlock{funcscalet}{\textbf{Scale in $t$:}}{}

\SetKwInput{KwIn}{Input}
\SetKwInput{KwDerIn}{Derived input}
\SetKwInput{KwOut}{Output}

\let\oldnl\nl% Store \nl in \oldnl
\newcommand{\nonl}{\renewcommand{\nl}{\let\nl\oldnl}}% use command \nonl to remove line number for one line

\begin{algorithm}%[H]
\caption{Scaling of velocity fields from \textit{base} to \textit{target} $\Rey$}
\KwIn{$u_{i}(y, z, t),~\textit{target}~\Rey$}%~\Rey_{\text{target}}$}
\KwDerIn{$U_{i,~\text{target}}(y)$, $sc$, $E_{sc}$, $n_{\text{modes}}$}
\KwOut{$u_{i,~\text{target}}(y, z, t)$}

Perform DFT in $z$: $\hat{u_i}^{(k_z)}(y,t) \leftarrow \mathcal{F}_z[u_i(y, z, t)] / \sqrt{n_z/2}$\

\label{algo_whole}

\AllLoop{

    Compute POD via SVD: $\hat{u_i}^{(k_z)}(y,t) \approx \sum\limits^{n_{\rm modes}-1}_{n=0}a^{(k_z,n)}(t)\hat{\varphi_i}^{(k_z,n)}(y)$\ 
    }

Classify $k_z$ into inner and outer spectral ranges: $\mathcal{K}_{\text{inner}}$, $\mathcal{K}_{\text{outer}}$\

\OuterLoop{
    %Scale/shift in $z$: $\hat{\varphi_i}^{(\text{round}(k_z/sc),n)}(y) \leftarrow \hat{\varphi_i}^{(k_z,n)}(y)$\;

    \funcscalez{

    $\hat{\varphi_i}^{(\text{round}(k_z/sc),n)}(y) \leftarrow \hat{\varphi_i}^{(k_z,n)}(y)$\
    
    }
}

From here, $\mathcal{K}_{\text{outer}}$ corresponds to the new scaled range of outer $k_z$, $\emph{i.e., } \mathcal{K}_{\text{outer}}\approx \text{round}(\mathcal{K}_{\text{outer}}/sc)$  

Inverse DFT of each POD mode for all $k_z\in (\mathcal{K_{\text{inner}}} \cup~\mathcal{K_{\text{outer}}})$: 
$\varphi_i^{(k_z,n)}(y,z)\leftarrow \mathcal{F}^{-1}_z[\hat{\varphi_i}^{(k_z,n)}(y)]\cdot\sqrt{n_z/2}$

\OuterLoop{

        %Scale $y$: $y_{\text{new}} = y * sc$\;

    \funcscaley{
    Upscale $y$-axis: $y_{\text{target}} \leftarrow y \cdot sc$\
    
    Remove $y_{\text{target}}>\text{max}(y)$ such that max($y_{\text{target}}$) $\approx$ max($y$). Consequently, size($y_{\text{target}}$) $<$ size($y$) 
    (\emph{i.e.,} $y_{\text{target}}$ ends up as a coarser grid than $y$)\
    
    2-D interpolate POD modes from the upscaled, coarse $y_{\text{target}}-$grid onto the finer \textit{base} $y-$grid: ${\varphi_i}^{(k_z,n)}(y_{\text{target}},z) \rightarrow {\varphi_i}^{(k_z,n)}(y,z)$\
    }

    %Remove $y_{\text{target}}>\text{max}(y)$ such that: $y_{\text{target}}[1:\text{idx}] = y [every other]$\;    

    %Interpolate on finer $y$ modes: $\phi_i(y,k_z) \rightarrow \phi_i(y_{\text{new}},k_z)$\;

    %Scale time: $t_{\text{new}} = t / sc$\;

    %Interpolate time coefficients: $a_i(k_z,t) \rightarrow a_i(k_z,t_{\text{new}})$\;

    \funcscalet{
    $t_{\text{target}} \leftarrow t / sc$\

    1-D interpolate POD time coefficients onto the scaled time axis : $a^{(k_z,n)}(t) \rightarrow a^{(k_z,n)}(t_{\text{target}})$\
    }
}\
%Reconstruct velocity fields at \textit{target~\Rey}:
%$u_{i,~\text{target}}(y,z,t_{\text{target}}) \leftarrow \sum_{k_z\in \{0, \mathcal{K}_{\text{inner}}, \mathcal{K}_{\text{outer}} \}}  \sum\limits^{n_{\rm modes}-1}_{n=0}a^{(k_z,n)}(t_{\text{target}}){\varphi_i}^{(k_z,n)}(y)e^{2\pi i k_zz/L_z}$  \
\begin{flalign*}
&\text{Reconstruct velocity fields at \textit{target} } \Rey: &&\\
&u_{i,~\text{target}}(y,z,t_{\text{target}}) \leftarrow \quad U_{i,~\text{target}}(y) \quad \text{(for } k_z = 0,~n = 0\text{)} &&\\
&\quad + \sum_{k_z\in \mathcal{K}_{\text{inner}}} \sum_{n=0}^{n_{\text{modes}} - 1} a^{(k_z,n)}(t) \varphi_i^{(k_z,n)}(y,z) &&\\ %e^{2\pi i k_z z / L_z} &&\\
&\quad + \sum_{k_z\in \mathcal{K}_{\text{outer}}} \sum_{n=0}^{n_{\text{modes}} - 1} E_{sc} \cdot a^{(k_z,n)}(t_{\text{target}}) \varphi_i^{(k_z,n)}(y,z) &&\\ %e^{2\pi i k_z z / L_z} &&
\end{flalign*}\

\end{algorithm}

%%%

\section{Results}
\label{sec:results}

The proposed method is evaluated in two ways: 

\begin{enumerate}
    \item assessing the quality of the generated 2D velocity slices that will be used as inlet condition in \S\ref{sec:quality_slices}, and

    \item applying those velocity planes as inflow condition and analyzing the evolution of a turbulent boundary layer starting at $Re_\theta=8000$ in \S\ref{sec:apply_inflow}. 
    
\end{enumerate}

For this, from the available database~\citep{EitelAmor_2014}, time dependent velocity slices at $Re_\theta=2240$ and $Re_\theta=4430$ are scaled up to a \textit{target} $Re_\theta=8000$ using the proposed method, and compared against the data at $Re_\theta=8000$ available from the database. The first two sets of scaled velocity planes will now be referred to as \textit{re2k\_sc} and \textit{re4k\_sc}, whereas the reference planes (from~\citet{EitelAmor_2014}) at $Re_\theta=8000$ will be referred to as \textit{re8k} (although all three of them are in principle at $Re_\theta=8000$). %, either by evolution or by using the proposed scaling method). 
The scaled velocity planes are expected to have a gap in the pre-multiplied spectra (as will be discussed in \S\ref{sec:quality_slices} and shown in Figure~\ref{fig:allRe_lambda_z}), which needs to be filled up during adaptation. To examine if the scaled cases fill up this missing region in a reasonable way, we add an extra case for comparison by modifying the reference \textit{re8k} case in such a way that the spanwise wavenumbers approximately between $500<\lambda_z^+<1400$ are removed to create the \textit{re8k\_onlyIO} case (which stands for ``only inner and outer"  regions).
 
%\subsection{Spectra}
%\subsection{Main simulations}

\subsection{Quality of inflow velocity slices}
\label{sec:quality_slices}

\begin{figure}
    \centering
    \begin{subfigure}[b]{6cm}
        \centering                
        \includegraphics[width=6cm]{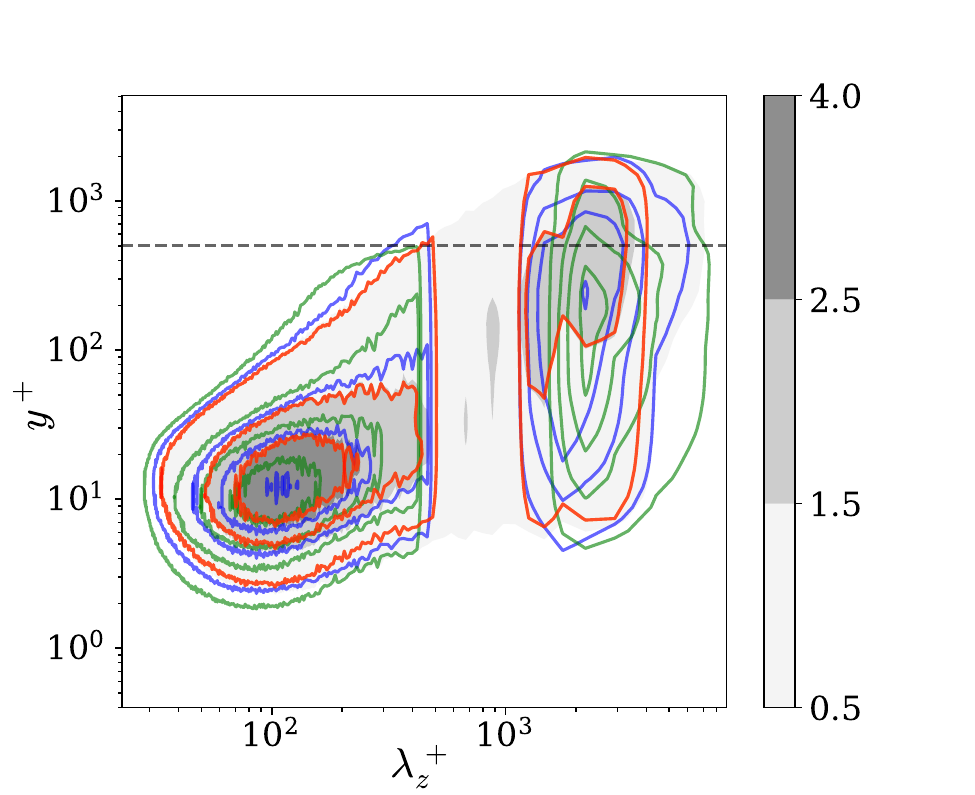}
        \caption{}
        \label{fig:allRe_lambda_z}
    \end{subfigure}
    \begin{subfigure}[b]{6cm}
        \centering
        \includegraphics[width=6cm]{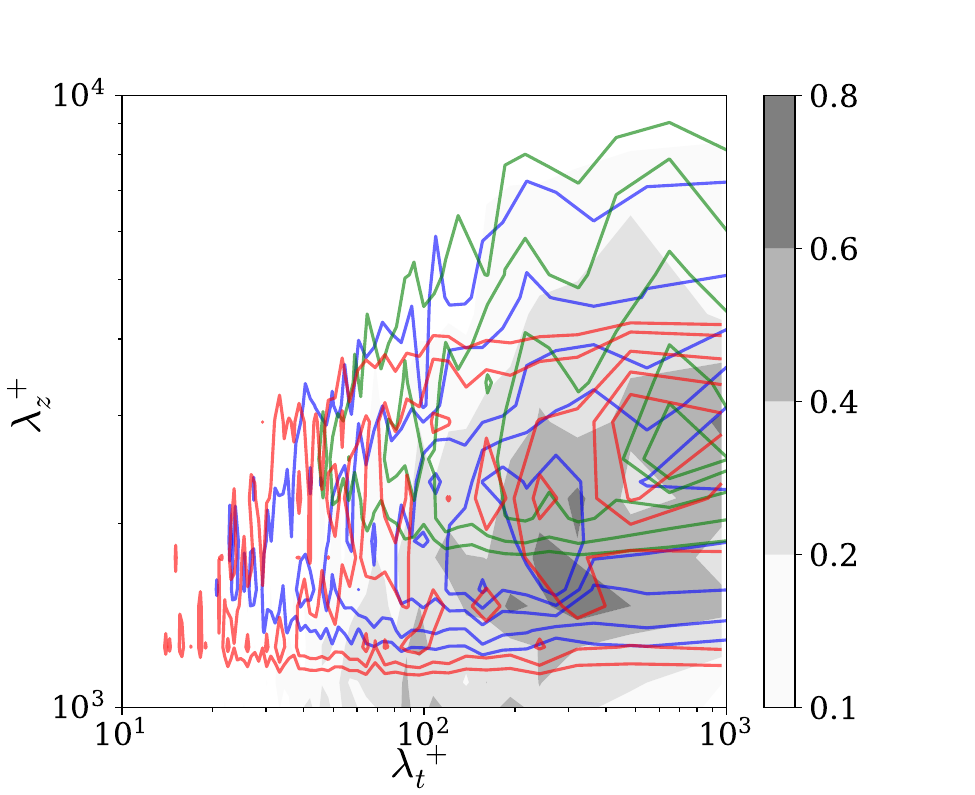}
        \caption{}
        \label{fig:allRe_lambda_zt}
    \end{subfigure}
    \caption{(a) 1D pre-multiplied spectra of streamwise velocity $u$, \emph{i.e.,} $E_{uu}(k_z)\cdot k_z /u_\tau^2$, and (b) 2D pre-muliplied PSD of $u$ in terms of $\lambda_z^+$ and $\lambda_t^+$, \emph{i.e.,} $E_{uu}(k_t k_z)\cdot k_t k_z /u_\tau^2$, at $y\approx 0.2\delta_{99_0}$ (this height is marked using a dashed horizontal line in (a)). In both figures, the reference data \textit{re8k} is shown in the background as filled contours using shades of gray with levels as shown in the color bar. \textit{re2k\_sc}, \textit{re4k\_sc} and \textit{re8k\_onlyIO} are shown using green, blue, and red contour lines respectively. All three showing the same contour levels as \textit{re8k}, but without using any shades of the respective color. $^{+}$ indicates normalization using inner units at the \textit{target} $\Rey$.}
    \label{fig:allRe_lambda_z_lambda_zt}
\end{figure}

%\begin{figure}
%    \centering
%    \includegraphics[width=0.5\linewidth]{./figures/re_stress_inlet_all.pdf}
    %\includegraphics[width=12cm]{./figures/Graphical_abstract.pdf}
%    \caption{Reynolds stress profiles of the inlet velocity slices. All curves shown here  have been passed through a median filter, without altering the general trend, to smoothen the noisy parts resulting from the coarse $y$-spacing in the precursor data. $^{+}$ indicates normalization using inner units.}
%    \label{fig:re_stress_inlet_all}
%\end{figure} 

\begin{figure}
    \centering
    \begin{subfigure}[b]{6cm}
        \centering                
        \includegraphics[width=6cm]{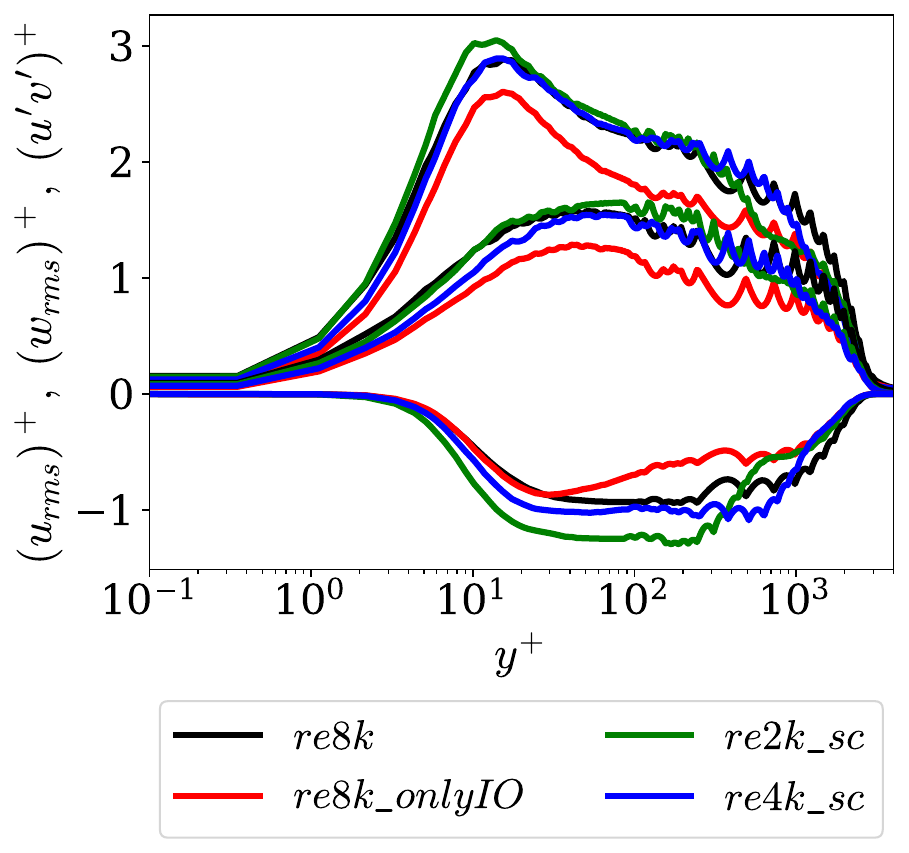}
        \caption{}
        \label{fig:re_stress_inlet_all_wavy}
    \end{subfigure}
    \begin{subfigure}[b]{6cm}
        \centering
        \includegraphics[width=6cm]{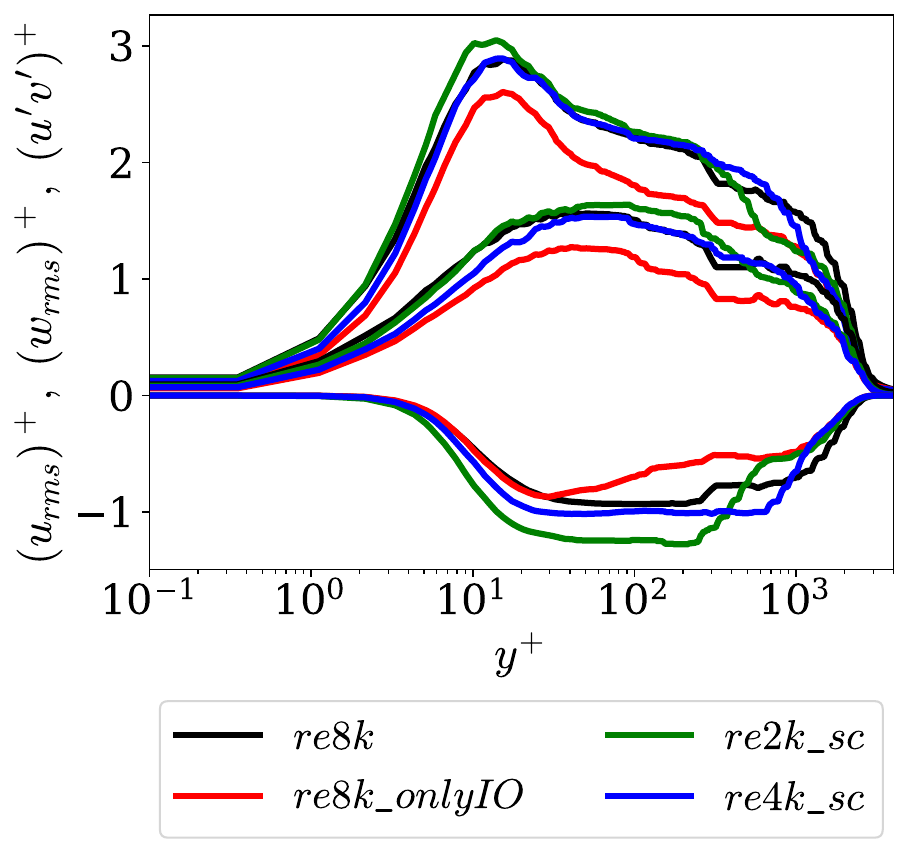}
        \caption{}
        \label{fig:re_stress_inlet_all_nonwavy}
    \end{subfigure}
    \caption{Reynolds stress profiles of the inlet velocity slices. (a) is the raw data whereas, (b) is obtained after passing it through a median filter, without altering the general trend, to smoothen the noisy parts to facilitate easier visual inspection here. The noise results from the coarse $y$-sampling in the precursor data. $^{+}$ indicates normalization using inner units at the inlet.}
    \label{fig:re_stress_inlet_all}
\end{figure}

The newly generated time-dependent, 2D velocity slices of the cases \textit{re2k\_sc} and \textit{re4k\_sc} at the \textit{target} $Re_\theta=8000$ are assessed here in comparison to \textit{re8k} and \textit{re8k\_onlyIO}, before applying it as an inflow condition in \S\ref{sec:apply_inflow}. 

Figure~\ref{fig:allRe_lambda_z} shows 1D pre-mulitplied spectra of streamwise velocity of the different cases. The gap in the spectra of the scaled cases (\emph{i.e.,} \textit{re2k\_sc} and \textit{re4k\_sc}), which arises due to the scaling procedure in $k_z$ (as described in \S\ref{sec:shifting_in_kz}), as well as the intentional gap in \textit{re8k\_onlyIO} is visible in the approximate range of $500<\lambda_z^+<1400$. Regardless, the inner and outer regions are reproduced quite reasonably well. In order to examine if the temporal evolution of the scaled outer structures are reproduced in an acceptable manner, 2D pre-multiplied spectra (in span and time) are computed at a height $y\approx0.2\delta_{99_0}$, as shown in Figure~\ref{fig:allRe_lambda_zt}. Despite the apparent noise due to the limited time series, all cases reproduce the strong outer peak in a reasonably similar position on the spectra.
%which lies towards the top right corner. %\textit{re8k\_onlyIO} also shows some weaker contributions towards the bottom left corner which is absent in the \textit{re2k\_sc} and \textit{re4k\_sc} cases due to the fact that the inner region is not scaled in our procedure and we expect it to be recovered rather rapidly.
These comparisons have now revealed that the scaling procedure in space and time works sufficiently well to proceed with the method.

In order to examine the effect of scaling of energy at the inlet plane, we look at the Reynolds stress profiles in Figure~\ref{fig:re_stress_inlet_all}. Due to the fact that the precursor database had time series data sampled on a coarse $y$-grid of about 47 points, these profiles are noisy towards the outer region. Nevertheless, that does not affect the proposed method. Owing to the way in which the scaling of energy is performed, as mentioned in~\S\ref{sec:method_scaling_energy}, where the square of the scaling factor is found by comparing the mean of the streamwise fluctuations beyond $y^+>100$ between the \textit{base} and \textit{target} profiles, followed by applying the scaling factor to a limited set of outer structures that are introduced into the \textit{target} plane (\emph{i.e.,} referring to the gap in the pre-multiplied spectra, see Figure~\ref{fig:allRe_lambda_z}); the scaled profiles may have a slight over-prediction of energy through out the profile and not just far away from the wall (due to the wall-attached nature of the large structures that are scaled) at the inlet. This explains why the scaled cases have higher energy than the \textit{re8k\_onlyIO} case, which has approximately similar spanwise structures missing as seen on the pre-multiplied spectra, but has the right amount of fluctuations in those retained structures as compared to \textit{re8k}. The difference in energy content between the \textit{re8k} reference case and the \textit{re8k\_onlyIO} case is caused entirely due to the missing spanwise wavenumbers in the pre-multiplied spectra. Looking at the region (\emph{i.e.,} approximately $100<y^+<2000$) where scaling of energy is applied, one sees how the $u_{rms}$ and $w_{rms}$ profiles of the scaled cases lift up from the $re8k\_onlyIO$ curve and overlaps the \textit{re8k} curve - showing that scaled outer structures are slightly over-stimulated to account for the energy of the missing structures. The $u'v'$ curve shows that the scaled cases slightly over-shoot the \textit{re8k} case, which potentially stems from the fact that the scaling factor in $v'v'$ may not be exactly the same as that of $u'u'$.

\subsection{Application as inflow condition}
\label{sec:apply_inflow}

% Serial parameters
\begin{table}%[h!]
\centering
%\begin{center}
\begin{tabular}{llll}
\textbf{Length} & \textbf{Grid points} & \textbf{GLL spacing} & \textbf{Element size}\\
&(Ele$\times$GLL)& (in inner units)& (in $\theta_0$)\\
&&&\\
 X=$100~\theta_0$& 400$\times$8=3200 & $\Delta~x^+=4.7-15.5$  &  0.25\\
 &&Avg = 10.13&\\
 &&& \\
 $Y_1$=$0-16~\theta_0$& 60$\times$8=480 & $\Delta~y^+=0.3-41$ & $0.02-0.6$ \\
~~~~($0-2~\delta_{99_0}$) && Avg = 20&Avg = 0.3\\  
  &&& \\
    $Y_2$=$16-40~\theta_0$ & 10$\times$8=80 & $\Delta~y^+\approx~19-277$ & $1.0-4.5$\\
    ~~~~($2-5~\delta_{99_0}$)& & Avg = 148  & Avg = 2.7\\    
&&& \\
 Z=$29.6~\theta_0$ & 225$\times$8=1800 & $\Delta~z^+=2.5-8.0$ & 0.13\\
 
 &&Avg = 5.3 &\\
 &&& \\
 \text{Total points:}& $(6.3\times10^6) \times 8^3$ & & \\
 (Ele$\times$GLL)&$=3.2\times10^9$&&\\
\end{tabular}   
%\end{center}
\caption{Details of the Neko DNS grid used for TBL simulations. $X,~Y,~Z$ are streamwise, wall-normal and spanwise directions respectively. `Ele' stands for the number of spectral elements used for spatial discretization, `GLL' stands for the number of Gauss--Lobatto--Legendre points used within each element and `Avg' stands for average. $^+$ denotes inner scaling based on the inlet mean velocity profile at $\Rey_\theta=8000$.}
\label{table:grid_plate_8000_5}
\end{table}

\subsubsection{Computational setup}
\label{sec:setup}

%\begin{figure}
%    \centering
%    \includegraphics[width=0.8\linewidth]{./figures/Schm_cases.pdf}
    %\includegraphics[width=12cm]{./figures/Graphical_abstract.pdf}
%    \caption{}
%    \label{fig:schematic_cases}
%\end{figure} 

The robustness and accuracy of the proposed method when used as an inflow condition for spatially developing turbulent boundary layers is examined here by using the velocity slices as a Dirichlet inlet condition in the CFD solver Neko~\citep{jansson2024neko}. Neko is a continuous Galerkin  spectral element code for the incompressible Navier Stokes equations. Velocity-pressure decoupling is performed using the $P_N-P_N$ formulation, time integration is third-order semi-implicit and dealiasing of the convective term is performed using the $3/2$-rule~\citep{Deville_Fischer_Mund_2002,nek5000_2021_tech_report}. A flat plate at $Re_\theta=8000$ is setup on a domain with $6.3\times10^6$ hexahedral elements using seventh-order Gauss--Lobatto--Legendre polynomials resulting in a total of $3.2\times10^9$ grid points with spacing as shown in Table \ref{table:grid_plate_8000_5}. In the wall-normal region denoted as $Y_1$, tanh spacing with a growth factor of 2.5 is used; whereas in the $Y_2$ region, geometric stretching with a stretching parameter of 1.18 is used. Average grid spacing in the outer region at the inlet is about $15~{\Delta~{y}}^+$, and that at the outlet is about $20~{\Delta~{y}}^+$ (where $^+$ denotes inner scaling is based on the inlet mean velocity profile at $\Rey_\theta=8000$).

In terms of the Dirichlet inflow condition, velocity slices are available either directly from the database~\citep{EitelAmor_2014} or from the scaling procedure (stemming indirectly from the same database), spaced in time about $0.044~\theta_0/U_\infty$ (or $0.47~\l_0^*/u_{\tau_0}$) apart. There are about 20 time steps in Neko in between two such available velocity fields, where a third-order Lagrange interpolation is performed to obtain inflow fields in between. The database contains a total of 19410 time instances, which corresponds to about $3.7~\delta_{99_0}/u_{\tau_0}$. From the flat plate TBL simulations, the first $0.88~\delta_{99_0}/u_{\tau_0}$ (2 flow through times) are removed to account for initial transients, and time- and span-averaging is performed for the next $2.8~\delta_{99_0}/u_{\tau_0}$ ($\approx 6.4$ flow through times). To ensure temporal convergence, one case was run for 3 more flow through times and no significant change was observed. Turbulent statistics are collected in Neko in a similar way as done by~\citet{kth_framework-2024}. Some parts of post-processing were performed using PySEMTools~\citep{perez2025_pysemtools}. Time series data is also collected from two cross-stream planes of probes located at two downstream locations ($x=50~\theta_0$ and $95~\theta_0$ corresponding to a location in the middle and towards the end of the domain of total length $100~\theta_0$) for the same duration as that of collection of statistics. In terms of the other boundary conditions, outflow condition is applied at the streamwise end of the domain, no-slip condition is applied at the bottom wall and periodicity is enforced in the spanwise direction. At the top boundary, outflow is allowed in the normal direction and a full-slip condition is used for the wall-parallel components of velocity. 

\begin{figure}
    \centering
    \begin{subfigure}[b]{6cm}
        \centering                
        \includegraphics[width=6cm]{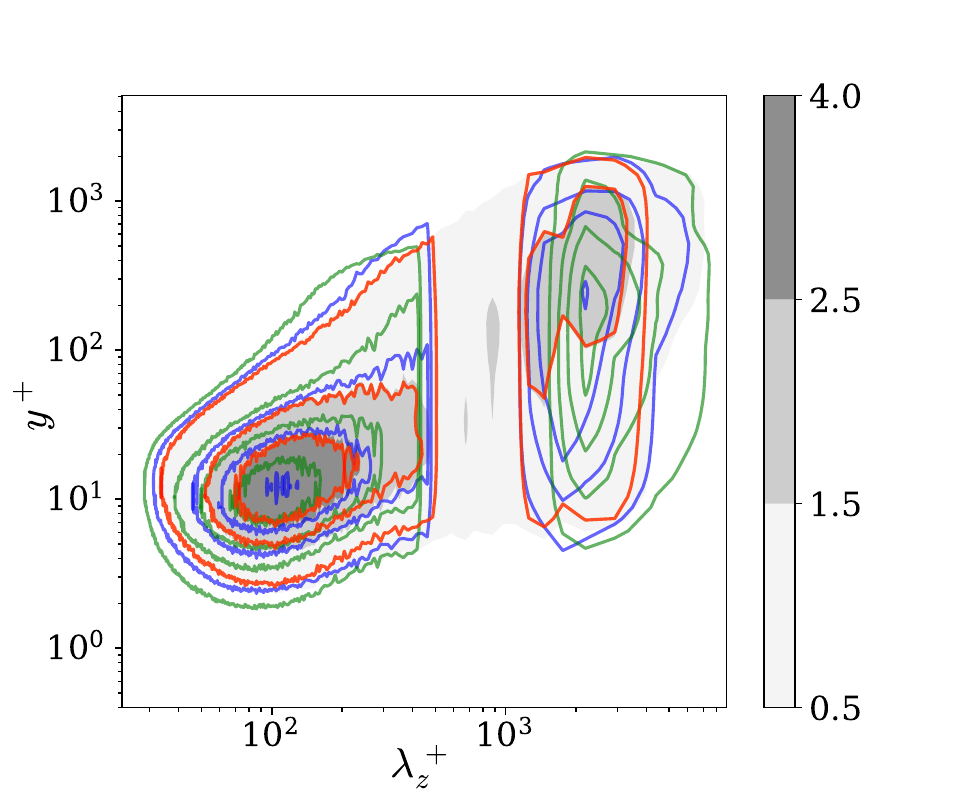}
        \caption{$x=0~\theta_0$ ; $Re_\theta\approx8000$}
        \label{fig:allRe_lambda_z_x0}
    \end{subfigure}
    \begin{subfigure}[b]{6cm}
        \centering                
        \includegraphics[width=6cm]{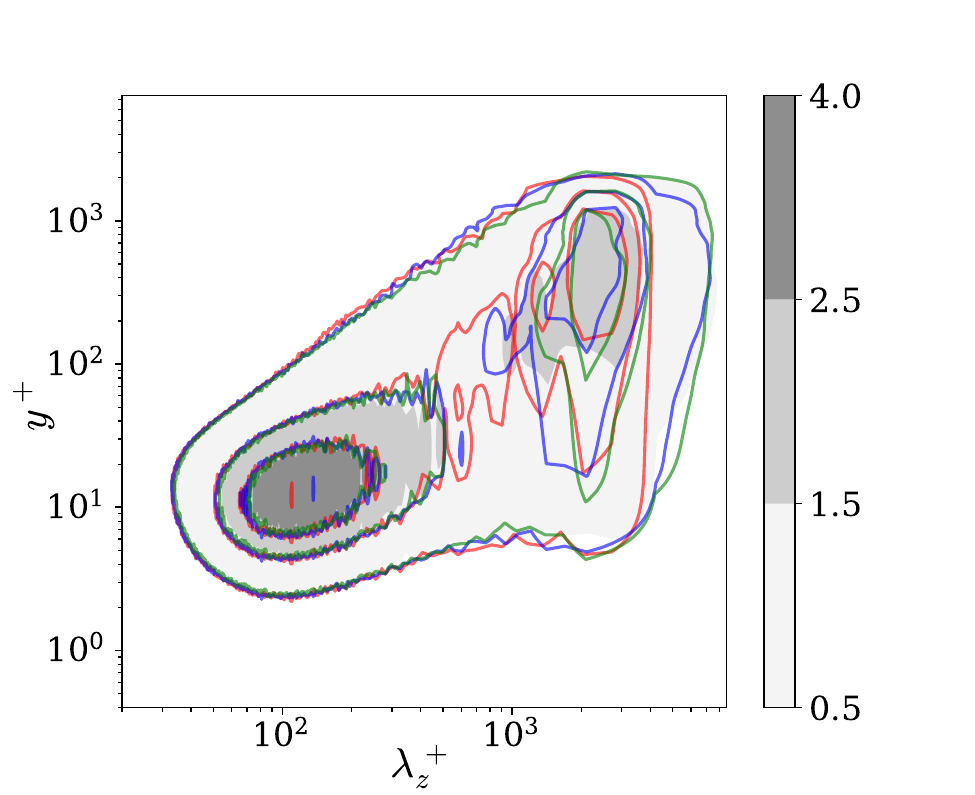}
        \caption{$x=50~\theta_0$ ; $Re_\theta\approx8500$}
        \label{fig:allRe_lambda_z_x50}
    \end{subfigure}
    \begin{subfigure}[b]{6cm}
        \centering
        \includegraphics[width=6cm]{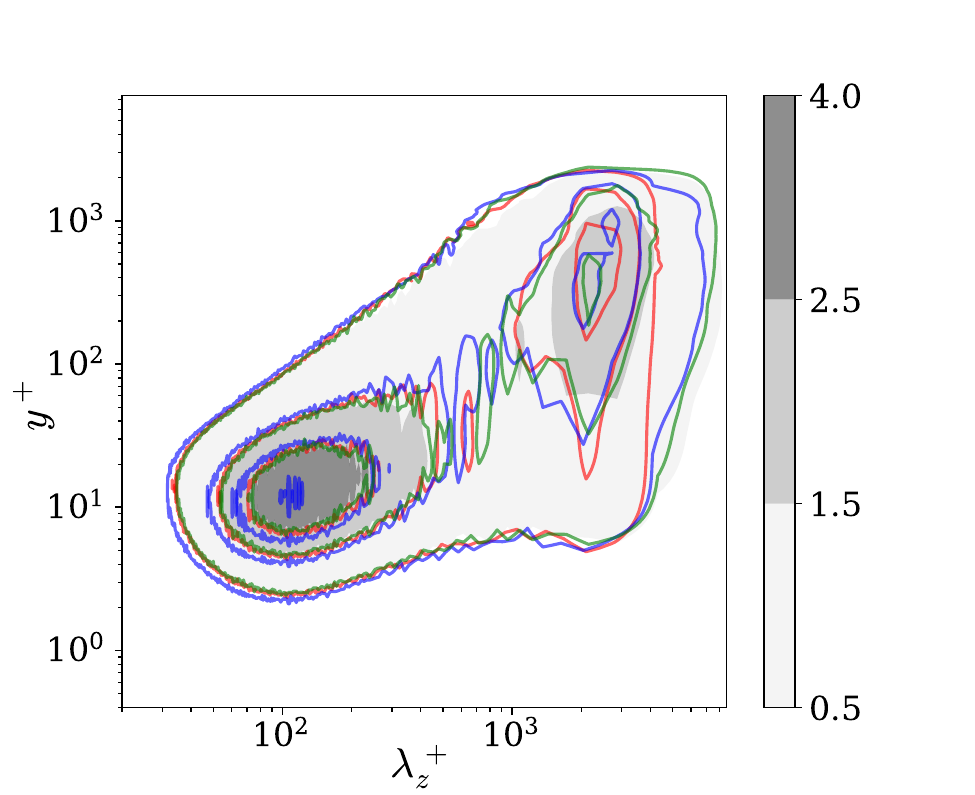}
        \caption{$x=95~\theta_0$ ; $Re_\theta\approx9000$}
        \label{fig:allRe_lambda_z_x95}
    \end{subfigure}
    \caption{1D pre-multiplied spectra of streamwise velocity $u$, \emph{i.e.,} $E_{uu}(k_z)\cdot k_z /u_\tau^2$, at streamwise positions (a) $x=0~\theta_0$ , (b) $x=50~\theta_0$ and (c) $x=95~\theta_0$. In all three figures, the reference data \textit{re8k} is shown in the background as filled contours using shades of gray with levels as shown in the color bar. \textit{re2k\_sc}, \textit{re4k\_sc} and \textit{re8k\_onlyIO} are shown using green, blue, and red contour lines respectively. All three showing the same contour levels as \textit{re8k}, but without using any shades of the respective color. $^{+}$ indicates normalization using inner units at the respective $\Rey$.}
    \label{fig:allRe_lambda_z_x0_50_95}
\end{figure}

\subsubsection{Development of the turbulent boundary layer}

Here the turbulent boundary layer produced by the proposed method is evaluated by examining its downstream development. Since the proposed method introduces large-scale structures that are absent in their corresponding \textit{base} case, and since the scaled planes lack certain spanwise wavenumbers (as evident from the gap in the pre-multiplied spectra in Figure~\ref{fig:allRe_lambda_z}), two main questions arise. Firstly, whether the newly introduced large-scale structures will fade away or continue to grow downstream, and secondly, whether and how fast can the gap in the spectra be filled in. 

%%% re stresses
\begin{figure}
    \centering
    \begin{subfigure}[b]{6cm}
        \centering                
        \includegraphics[width=6cm]{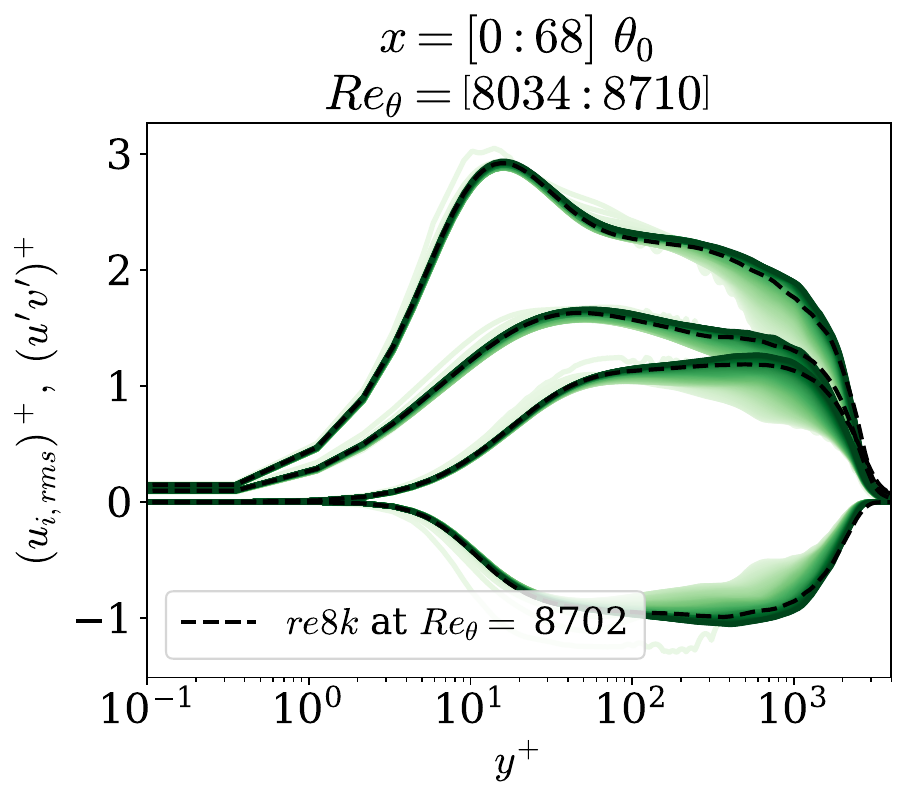}
        \caption{\textit{re2k\_sc}}
        \label{fig:re_stress_2ksc}
    \end{subfigure}
    \begin{subfigure}[b]{6cm}
        \centering
        \includegraphics[width=6cm]{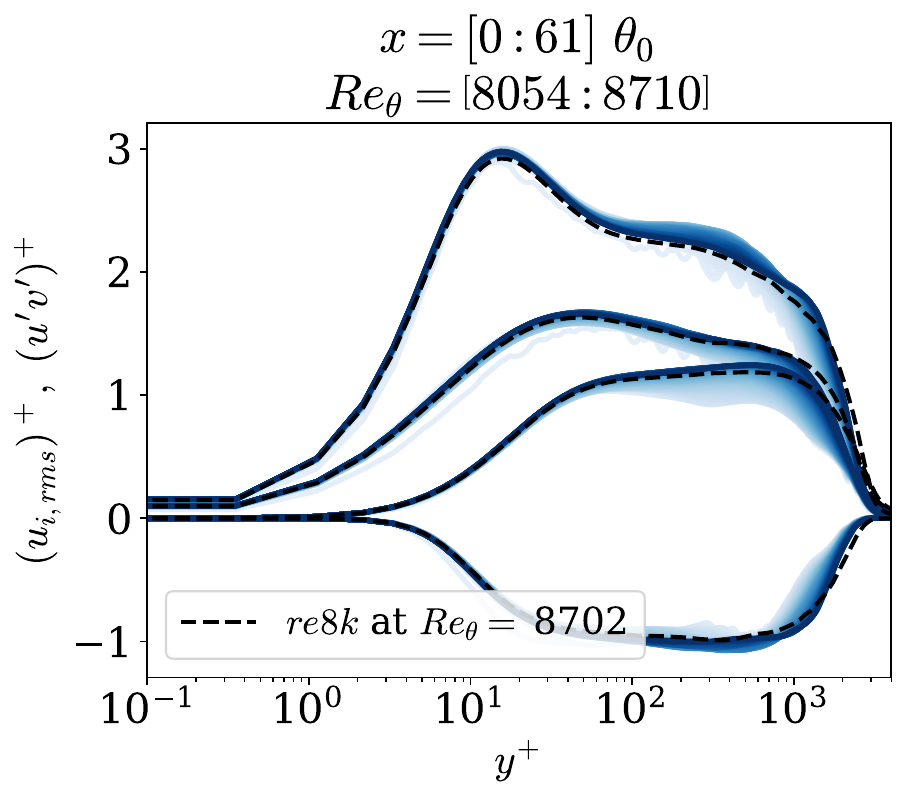}
        \caption{\textit{re4k\_sc}}
        \label{fig:re_stress_re4ksc}
    \end{subfigure}
    \begin{subfigure}[b]{6cm}
        \centering
        \includegraphics[width=6cm]{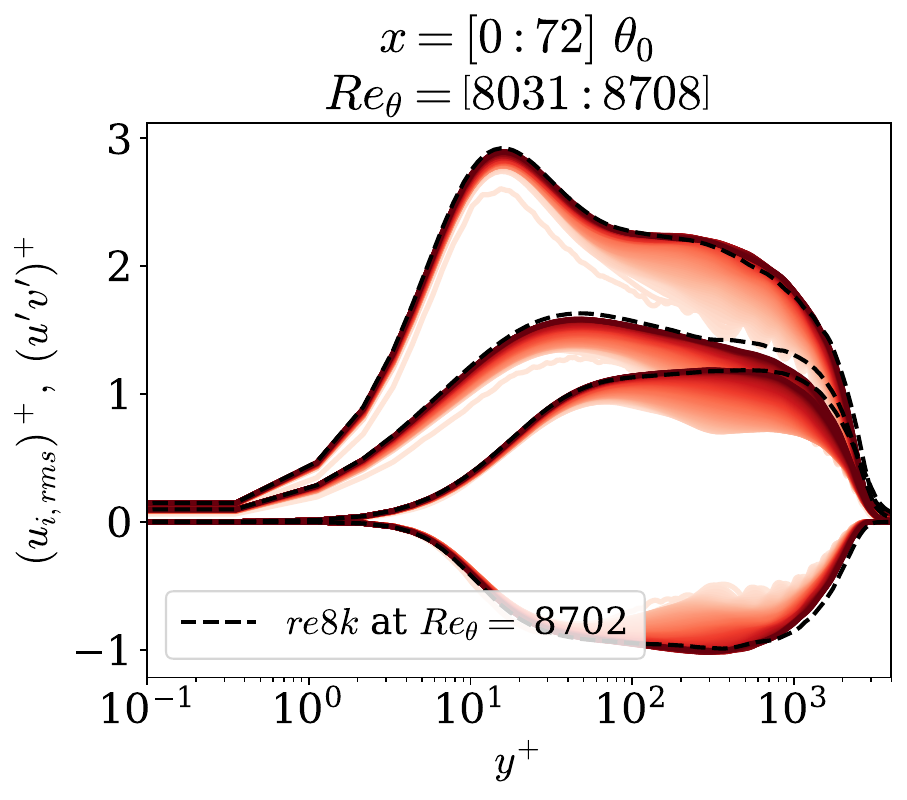}
        \caption{\textit{re8k\_onlyIO}}
        \label{fig:re_stress_re8kIO}
    \end{subfigure}
    \begin{subfigure}[b]{6cm}
        \centering
        \includegraphics[width=6cm]{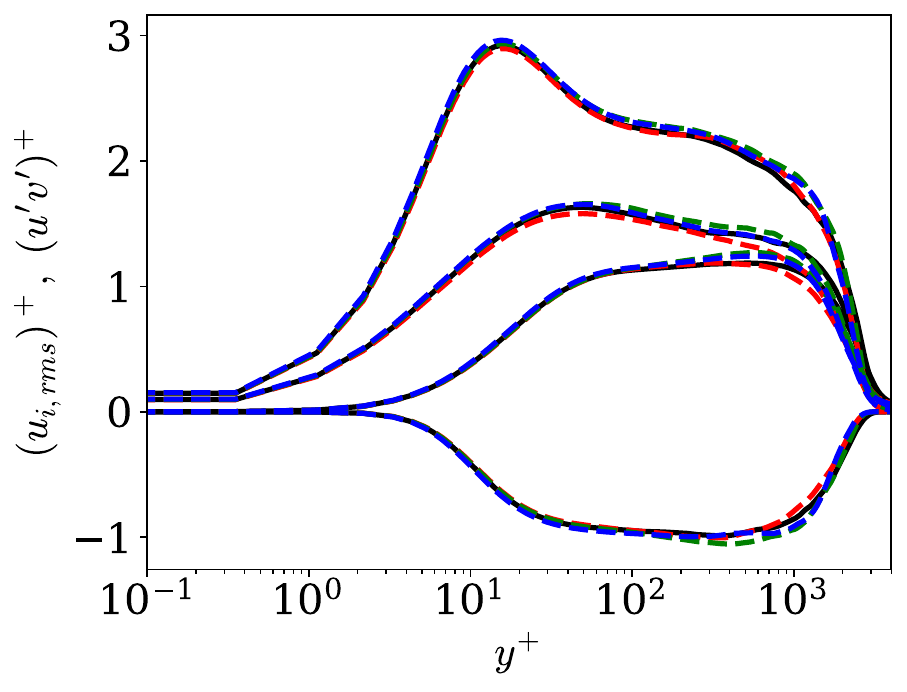}
        \caption{All cases at $Re_\theta\approx8700$}
        \label{fig:re_stress_re8700_all}
    \end{subfigure}    
    \caption{Streamwise development of Reynolds stress profiles for the different cases upto $Re_\theta=8700$ as compared to \textit{re8k}. Light to darker shades show increasing streamwise distance. $^{+}$ indicates normalization using inner units at the inlet.}
    \label{fig:re_stresses_all}
\end{figure}
%%%

To answer these questions, we start with the time series data extracted at the two downstream locations and examine the pre-multiplied spectra of streamwise velocity at these locations in comparison to the same at the inlet, as shown in Figure~\ref{fig:allRe_lambda_z_x0_50_95}. It is straight away clear that the large-scale structures have survived through-out the domain, as well as the gap getting filled up presumably well before the middle of the domain where we have the probes placed (\emph{i.e.,} $x=50~\theta_0\approx 6~\delta_{99_0}$). Interestingly, the scaled cases filled up faster and better matches the reference \textit{re8k} case than the \textit{re8k\_onlyIO} case. Looking at the largest wavelengths, it is clear how scaling the energy helped the scaled cases to closely match the shape of the spectra of the large-scale structures of the reference \textit{re8k} case as opposed to \textit{re8k\_onlyIO}. The slight deviations in the inner region between the scaled cases and the \textit{re8k} at the inlet, due to not having scaled the inner region, has also long disappeared by the time the flow reaches the middle of the domain.

In order to get a better picture of how the turbulent fluctuations evolve downstream the inlet, we look at the streamwise evolution of the Reynolds stress profiles in Figure~\ref{fig:re_stresses_all}. Here, the profiles starting from the inlet are plotted up to a streamwise distance (or equivalent $\Rey$) where most cases seemed to match \textit{re8k} at a corresponding $\Rey$ the earliest. After preliminary observations, this is chosen to be when the cases reach $\Rey_\theta\approx8700$. Since we start from a coarsely sampled data (as mentioned in \S\ref{sec:quality_slices} and shown in Figure~\ref{fig:re_stress_inlet_all}), it takes the first $12~\theta_0\approx1.4~\delta_{99_0}$ distance downstream the inlet for these Reynolds stress profiles to be smooth. Thereafter, \textit{re4k\_sc} reaches $\Rey_\theta\approx8700$ earliest at a downstream location of $x=61~\theta_0\approx7.3~\delta_{99_0}$ from the inlet, followed by \textit{re2k\_sc} at $x=68~\theta_0\approx8.1~\delta_{99_0}$ and lastly \textit{re8k\_onlyIO} at $x=72~\theta_0\approx8.6~\delta_{99_0}$. At that position, as seen in Figure~\ref{fig:re_stress_re8700_all}, \textit{re2k\_sc} and \textit{re4k\_sc} matches the profiles of \textit{re8k} much better than that of \textit{re8k\_onlyIO}. \textit{re8k\_onlyIO} under-predicts all the Reynolds stresses. Provided that the two scaled cases had \textit{base} $\Rey$ that were two times smaller and still gave great agreement for Reynolds stresses at about the same downstream distance of $\approx~8~\delta_{99_0}$, hints at a \textit{base} $\Rey$ independent development length for the proposed method. %, as long as outer scaling is valid. 

%cf_H12
\begin{figure}
    \centering
    \begin{subfigure}[b]{6cm}
        \centering                
        \includegraphics[width=6cm]{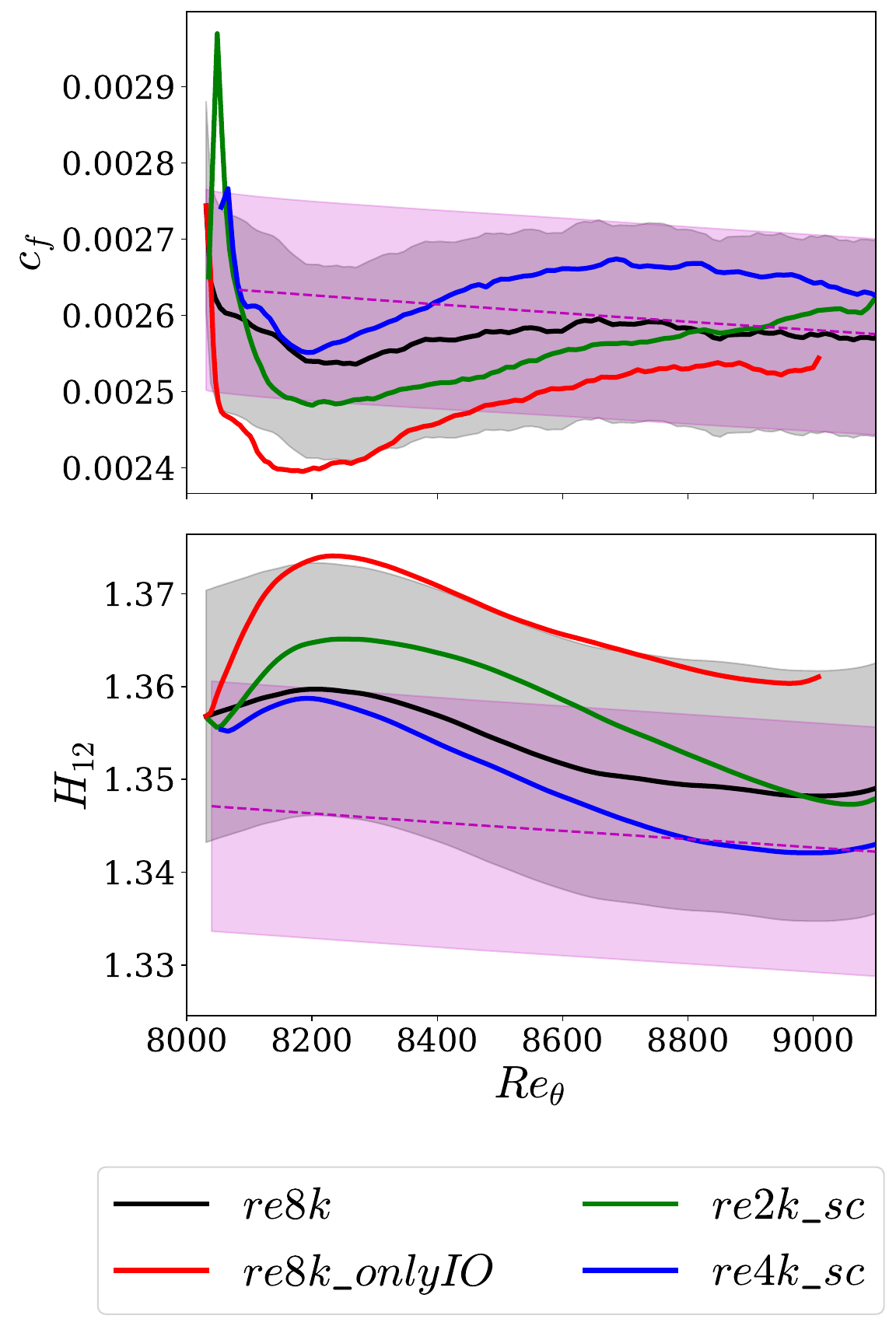}
        \caption{}
        \label{fig:cf_H12_vs_re}
    \end{subfigure}
%    \begin{subfigure}[b]{4cm}
%        \centering
%        \includegraphics[width=4cm]{./figures/cf_H12_vs_d99_0.pdf}
%        \caption{}
%        \label{fig:H12}
%    \end{subfigure}
    \begin{subfigure}[b]{6cm}
        \centering
        \includegraphics[width=6cm]{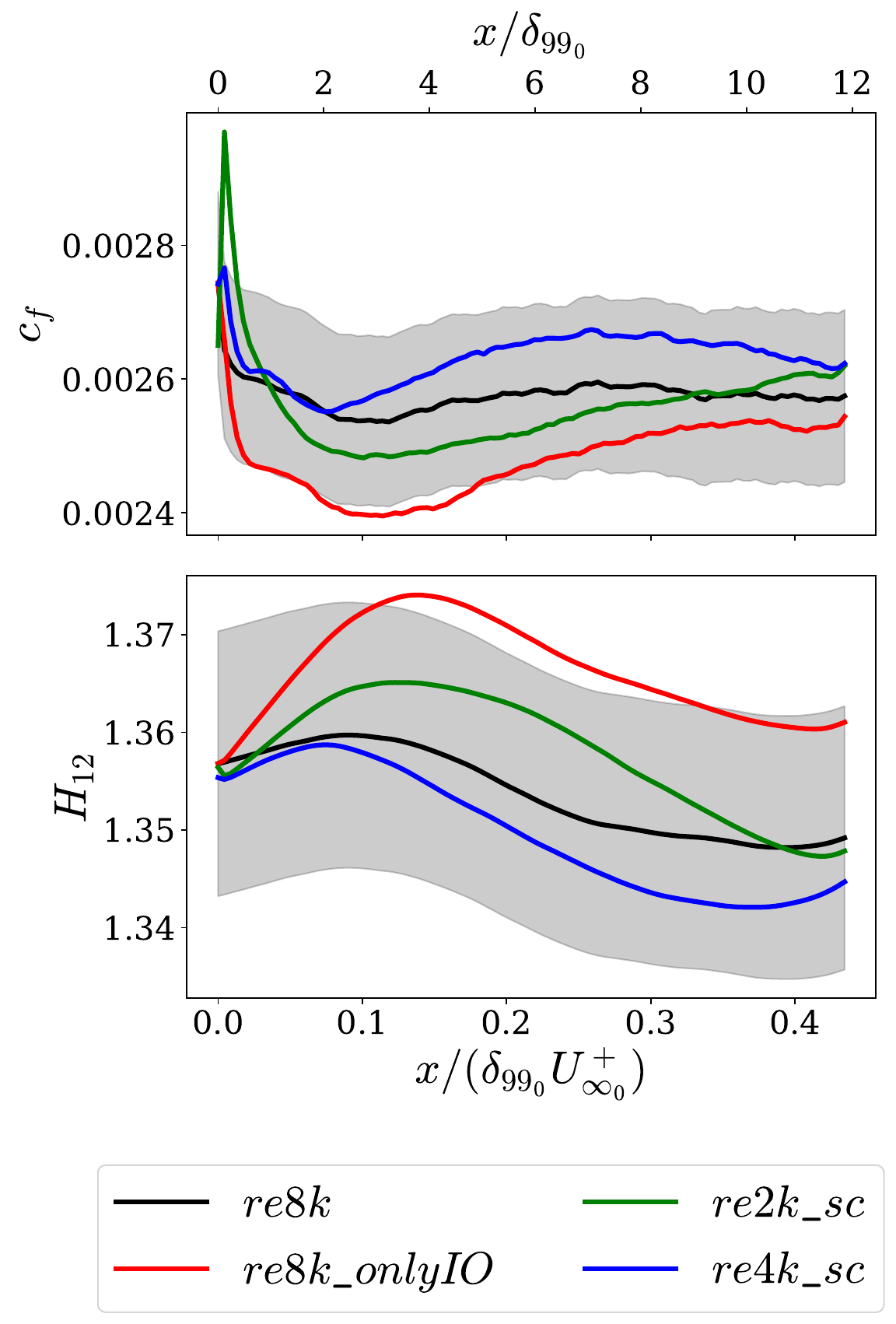}
        \caption{}
        \label{fig:cf_H12_vs_x}
    \end{subfigure} 
    \begin{subfigure}[b]{0.798\linewidth}
        \centering
        \includegraphics[width=0.798\linewidth]{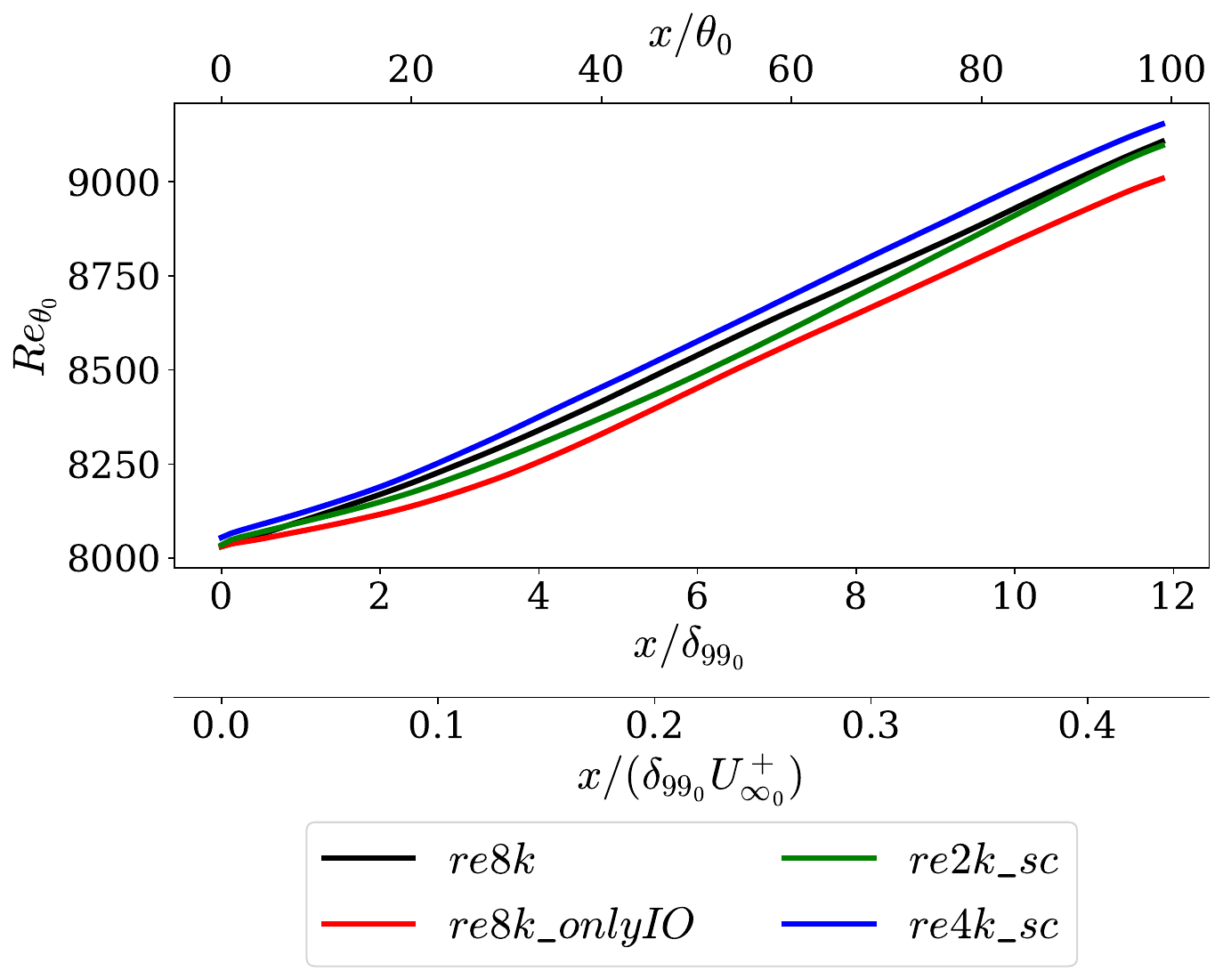}
        \caption{}   
        \label{fig:retheta_vs_x}
    \end{subfigure}
    \caption{Development of skin friction coefficient, $c_f$, and shape factor, $H_{12}$ plotted against (a) $\Rey_\theta$ and (b) non-dimensional streamwise distance. In (a), dashed line in magenta color is the Coles-Fernholz relation and correlation both by~\citet{Chauhan_2009}, for $c_f$ and $H_{12}$ respectively. Shaded region in magenta color is $\pm5\%$ and $\pm1\%$ tolerances with respect to the corresponding dashed magenta lines from these correlations for $c_f$ and $H_{12}$ respectively. In both (a) and (b), the grey shaded region is the same amount of tolerance but with respect to \textit{re8k}. (c) Development of $\Rey_\theta$ against different non-dimensionalized streamwise distances. The maximum spread in $\Rey_\theta$ at the inlet is $0.3\%$ and that at the outlet is $1\%$, both with respect to \textit{re8k}.}
    \label{fig:cf_H12}
\end{figure}

In Figure~\ref{fig:cf_H12}, the development length is further examined using the usual measures of inner- and outer-scale convergence~\citep{schlatter_örlü_2012,Chauhan_2009}, skin friction coefficient, $c_f$, and shape factor, $H_{12}=\delta^*/\theta$ (where $\delta^*$ is displacement thickness and $\theta$ is momentum-loss thickness), respectively. The development is shown against $\Rey_\theta$ in Figure~\ref{fig:cf_H12_vs_re}, and against two non-dimensionalised streamwise distances in Figure~\ref{fig:cf_H12_vs_x}; which can be examined along with Figure~\ref{fig:retheta_vs_x} where $\Rey_\theta$ is plotted against different non-dimensionalized streamwise distances. As a side note, it is worth mentioning that here $\theta$ and $H_{12}$ are calculated by integrating up to the top of the domain and taking $U_\infty=1$. One of those non-dimensional distances shown in Figures~\ref{fig:cf_H12_vs_x} and \ref{fig:retheta_vs_x} is the familiar length in terms of the inlet boundary layer thickness, $\delta_{99_0}$. Another one is based on large eddy turn-over length $\delta_{99} U_\infty^+$, following the suggestion by ~\citet{Simens_Jimenez2009} and used by~\citet{sillero_2013}. It characterizes the development length based on how far the eddies advect in a large-eddy turnover time $\delta_{99}/u_\tau$~\citep{sillero_2013}. Although here we use $\delta_{99_0} U_{\infty_0}^+$ instead, it is expected to be reasonably similar to $\delta_{99} U_\infty^+$ owing to the short domain where the extents of the TBL in terms of $\Rey_\theta$ is short, as well as the variations in $\delta_{99}$ and $u_{\tau}$ being small compared to the variation in $x$ along the domain. For instance, for \textit{re8k}, $\delta_{99}$ grows only by a factor $1.13$ from the inlet to the outlet of the domain. 

We begin by discussing the development of $c_f$ in comparison to the Coles--Fernholz relation, and $H_{12}$ against a correlation, both by~\citet{Chauhan_2009}. Since there is usually $\pm5\%$ and $\pm1\%$ spread between DNS and experimentally measured $c_f$ and $H_{12}$ respectively, with respect to these correlations~\citep{EitelAmor_2014, sillero_2013}, such tolerances are shown in Figure~\ref{fig:cf_H12_vs_re}. $c_f$ of both the scaled cases, \textit{re2k\_sc} and \textit{re4k\_sc}, as well as that of the precursor reference simulation, \textit{re8k}, stay within the $\pm5\%$ margin with respect to the correlations right from the inlet. It is only the \textit{re8k\_onlyIO} case that briefly drifts outside this tolerance shortly after the inlet. In terms of $H_{12}$, both the reference \textit{re8k} as well as the scaled \textit{re4k\_sc} case stays within these acceptable margins with respect to the correlations through-out the domain straight away from the inlet. The $H_{12}$ curve of \textit{re2k\_sc} comes into this region at $\Rey_\theta\approx8600$ ($\approx7.5~\delta_{99_0}$ or $\approx0.28~\delta_{99_0}U_{\infty_0}^+$). On the other hand, the \textit{re8k\_onlyIO} always stays outside these margins through out the domain. 

The development length of \textit{re8k} until $\Rey_\theta\approx8600$, (\emph{i.e.,} $\approx 6.5~\delta_{99_0}$ or $\approx 0.26~\delta_{99_0}U_{\infty_0}^+$, where its $c_f$ and $H_{12}$ comes as close to the correlation as possible and continues that trajectory; where it is exactly on top of the $c_f$ from Coles-Fernholz and $0.5\%$ away for $H_{12}$), is due to the coarse $y-$spacing of the inlet data as well as the time interpolation performed between two subsequent inlet velocity planes. As such these sources of errors are present in all cases we study here, and would vanish if inlet data at \textit{base} $\Rey$ that is more continuous in space and time is available. That contribution to development length is independent of the proposed inflow method. This necessitates a comparison with similar margins with respect to the \textit{re8k} case. All cases, also including \textit{re8k\_onlyIO}, always stay within the $\pm5\%$ and $\pm1\%$ tolerances with respect to the $c_f$ and $H_{12}$ respectively of \textit{re8k}. In fact, the scaled cases are always within $\pm3.5\%$ and $\pm0.5\%$ with respect to \textit{re8k} for these two quantities right from the inlet, which is uncommon with inflow methods that are not precursors. 

Considering all the above mentioned observations for the scaled cases, including the developments of Reynolds stresses in Figure~\ref{fig:re_stresses_all}, and $c_f$ and $H_{12}$ in Figure~\ref{fig:cf_H12}, we conclude that the new scaling procedure-based inflow generation method results in a development length largely determined by the development of Reynolds stresses. This development length is about $8~\delta_{99_0}$ or $0.3~\delta_{99_0}U_{\infty_0}^+$, for the cases studied here. This is at least one order of magnitude shorter than other high $\Rey$ inflow studies in literature. Notably, in the work by~\citet{sillero_2013}, where they performed DNS of a TBL up to $\Rey_\theta=6000$ and where the development length was dominated by the adaptation of large-scale structures, a development length of $3-4~\delta_{99}U_{\infty}^+$ was needed for $H_{12}$ to reach within $\approx\pm1\%$ tolerance of an empirical fit. And based on~\citet{Simens_Jimenez2009}, relaxation of most flow scales need at least $1~\delta_{99}U_{\infty}^+$, which is not true for the current method. In the proposed method, we avoid having to wait for the adaptation of the time consuming large-scale structures by introducing their physically accurate description, as well as retaining the near-wall structures. As such the short development length is then determined by how the region in the middle of the inner and outer regions in the spanwise pre-multiplied spectra is filled. This can be monitored by following the streamwise development of the Reynolds stress profiles.

\section{Conclusions and outlook}
\label{sec:discuss}

We presented an inflow generation method which is based on up-scaling a set of two-dimensional (2D) precursor velocity slices in spectral space from a lower \textit{base} $\Rey$ to higher \textit{target} $\Rey$ to be used as inflow condition for spatially developing TBLs. Apart from the precursor data at the lower \textit{base} $\Rey$ (which could be taken from an existing database, if the width matches the $\textit{target}$ domain, such as the one we use here), this method only needs the value of the \textit{target} $\Rey$ as inputs. 

The essence of the method involve providing a rich description of the computationally expensive large-scale structures as well as the physically relevant near-wall structures. This helps it to evade the usual factor that results in large development lengths for high $\Rey$ TBLs - the development of slow evolving outer structures. As such, the proposed method resulted in a development length ($\approx8~\delta_{99_0}$) that is one order of magnitude shorter than other high $\Rey$ TBL DNS in literature. Since the usual measures, such as the development of $c_f$ and $H_{12}$, gave reasonable agreement with the reference precursor simulation right from the inlet, the development length of the proposed method is determined by how the missing fluctuations in the pre-multiplied spectra got filled up and is tracked using the streamwise evolution of the Reynolds stress profiles.

The method worked seamlessly when \textit{base} $\Rey_\theta=2240$ and $4430$ were used to produce \textit{target} $\Rey_\theta=8000$. Starting from precursor data at a $\textit{base}~\Rey$ high enough, such that outer-scaling holds true, is necessary to seamlessly obtain the correct pre-multiplied spectra at the \textit{target} $\Rey$ using the scaling procedure. Nevertheless, in the interest of further stress-testing the method, we ran it using data from a very low $\textit{base}~\Rey_\theta=790$. Scaling this, on the other hand, required a couple of iterations of trial and error to find the right scaling parameters for space and energy to correctly match the pre-multiplied spectra at the \textit{target} $\Rey$. If one has access to such a spectra at the \textit{target} $\Rey$, for instance from experimental measurements or obtained by applying the scaling procedure on one of the $\textit{base}~\Rey$ used in this work, then such a trail and error exercise is worthwhile in the interest of going further down in the precursor $\textit{base}~\Rey$. Going to lower $\textit{base}~\Rey$ could be helpful if the existing database cannot be used, for instance if the width of the $\textit{target}$ domain does not match with that of the database, which necessitates a new precursor to be run on a wider domain but at a lower $\textit{base}~\Rey$ such that it is computationally affordable. %that needs to be run at the \textit{base} $\Rey$ to fit the right width of domain for the $\textit{target}$ velocity fields. %Going to lower $\textit{base}~\Rey$ could be helpful if the existing database cannot be used, for instance due to the width of the $\textit{target}$ domain not matching with that of the database, necessitating a new precursor to be run at the $\textit{base}~\Rey$. 

When scaling the \textit{base} $\Rey$ velocity slices to \textit{target} $\Rey$, the spanwise width of the \textit{base} velocity fields should be wide enough such that it can fit a couple of the widest structures resulting after scaling to the higher \textit{target} $\Rey$. In order for this not to be a limiting factor for using the proposed method, the best way is to make sure to have a wide enough domain for the \textit{base} velocity fields. Work-arounds to artificially increase the spanwise width of existing data should be done carefully, if not avoided. This can, for instance, be done either using zero-padding or by performing a windowing operation before replicating in span, to avoid discontinuities. Directly replicating the data in span, taking advantage of the spanwise periodicity, would result in spectral artifacts like artificial discontinuities due to spectral leakage. 

It is necessary to have a long time series of velocity planes, long enough to study the statistics of the evolving turbulent boundary layer without introducing artificial spanwise periodicity by re-starting from the same set of planes to continue the time signal. One can avoid this by either running a very long precursor at the  \textit{base} $\Rey$, or use methods like the vector autoregression based time series extension method recently proposed by~\citet{Stanly_Du-JFM-2024}. Using this, one can first extend the \textit{base} $\Rey$ data to arbitrary lengths in time, and then input it (either as just time coefficients or as reconstructed velocity fields) to the proposed method to scale those to a higher \textit{target} $\Rey$. 

Although not exploited to its full potential in this work, the SVD performed in the wall-normal direction permits low-order modeling of the wall-normal dynamics. We kept the same $n_{\text{modes}}$ across all $k_z$ such that a minimum amount of TKE across all $k_z$ was maintained. It is possible to obtain an optimized $n_{\text{modes}}$ for each $k_z$ to enable further reduction in the order of the model while maintaining a required amount of TKE. This is possible as the contribution of different $k_z$ to TKE is different thereby enabling further truncation.

The proposed method can naturally be applied to other wall-bounded flows, like pipe and channels for instance, to provide efficient inflow conditions at higher Reynolds numbers with minimal initial transients. This method can also be used to provide inflow condition for high $\Rey$ TBLs in applications such as wind turbine farms where atmospheric boundary layer needs to be simulated. Also, applications in other cases such as jets and mixing layers may be considered, as long as an approximate scaling law of both mean and fluctuating profiles is available.

\textbf{Declaration of interests.} The authors report no conflict of interest.

\textbf{Acknowledgements.} %ADD Excellerat and Ceec, and computer time. Essentially the same as in the other JFM article. 
The authors gratefully acknowledge Geert Brethouwer and Shiyu Du for valuable discussions and constructive comments on the manuscript, and Ardeshir Hanifi for insightful discussions. The computations and data handling were enabled by resources provided by the National Academic Infrastructure for Supercomputing in Sweden (NAISS), partially funded by the Swedish Research Council through grant agreement no. 2022-06725. In addition, NAISS is acknowledged for awarding a project access to the LUMI supercomputer, owned by the EuroHPC Joint Undertaking and hosted by CSC (Finland) and the LUMI consortium. This project has received funding from the European High-Performance Computing Joint
Undertaking (JU) under grant agreement No 101092621. The JU receives support from the
European Union’s Horizon Europe research and innovation programme and Germany, Italy,
Slovenia, Spain, Sweden and France.

\bibliographystyle{jfm}
\bibliography{jfm_inflow_scaling}

@article{Deshpande2021,
title = {Data-driven enhancement of coherent structure-based models for predicting instantaneous wall turbulence},
journal = {International Journal of Heat and Fluid Flow},
volume = {92},
pages = {108879},
year = {2021},
issn = {0142-727X},
doi = {https://doi.org/10.1016/j.ijheatfluidflow.2021.108879},
url = {https://www.sciencedirect.com/science/article/pii/S0142727X21001090},
author = {R. Deshpande and C. M. {de Silva} and M. Lee and J. P. Monty and I. Marusic}
}

@article{coherent_structure-Robinson-1991,
author = {Robinson, S. K.},
title = {Coherent Motions in the Turbulent Boundary Layer},
journal = {Annual Review of Fluid Mechanics},
volume = {23},
number = {1},
pages = {601-639},
year = {1991},
doi = {10.1146/annurev.fl.23.010191.003125},
URL = { https://doi.org/10.1146/annurev.fl.23.010191.003125
}
}

@article{EitelAmor_2014,
title = "{Simulation and validation of a spatially evolving turbulent boundary layer up to ${\rm Re}_\theta=8300$}",
journal = {International Journal of Heat and Fluid Flow},
volume = {47},
pages = {57-69},
year = {2014},
issn = {0142-727X},
doi = {https://doi.org/10.1016/j.ijheatfluidflow.2014.02.006},
url = {https://www.sciencedirect.com/science/article/pii/S0142727X14000319},
author = {Eitel-Amor, G. and Örlü, R. and Schlatter, P.},
}

@article{Jarrin_2006,
title = {A synthetic-eddy-method for generating inflow conditions for large-eddy simulations},
journal = {International Journal of Heat and Fluid Flow},
volume = {27},
number = {4},
pages = {585-593},
year = {2006},
note = {Special Issue of The Fourth International Symposium on Turbulence and Shear Flow Phenomena - 2005},
issn = {0142-727X},
doi = {https://doi.org/10.1016/j.ijheatfluidflow.2006.02.006},
url = {https://www.sciencedirect.com/science/article/pii/S0142727X06000282},
author = {Jarrin, N. and Benhamadouche, S. and Laurence, D. and Prosser, R.},

}

@article{Lund_1998,
title = {Generation of Turbulent Inflow Data for Spatially-Developing Boundary Layer Simulations},
journal = {Journal of Computational Physics},
volume = {140},
number = {2},
pages = {233-258},
year = {1998},
issn = {0021-9991},
doi = {https://doi.org/10.1006/jcph.1998.5882},
url = {https://www.sciencedirect.com/science/article/pii/S002199919895882X},
author = {Lund, T.S. and Wu, X. and Squires, K.D.},

}

@article{schlatter_örlü_2012, title={Turbulent boundary layers at moderate Reynolds numbers: inflow length and tripping effects}, volume={710}, DOI={10.1017/jfm.2012.324}, journal={Journal of Fluid Mechanics}, publisher={Cambridge University Press}, author={Schlatter, P. and Örlü, R.}, year={2012}, pages={5–34}}

@article{spalart_1988, title="{Direct simulation of a turbulent boundary layer up to ${\rm R}_\theta = 1410$}", volume={187}, DOI={10.1017/S0022112088000345}, journal={Journal of Fluid Mechanics}, publisher={Cambridge University Press}, author={Spalart, P. R.}, year={1988}, pages={61–98}}

@article{Tabor_2010,
title = {Inlet conditions for large eddy simulation: A review},
journal = {Computers \& Fluids},
volume = {39},
number = {4},
pages = {553-567},
year = {2010},
issn = {0045-7930},
doi = {https://doi.org/10.1016/j.compfluid.2009.10.007},
author = {Tabor, G.R. and Baba-Ahmadi, M.H.},
}

@article{wu_2017,
author = {Wu, X.},
title = {Inflow Turbulence Generation Methods},
journal = {Annual Review of Fluid Mechanics},
volume = {49},
number = {1},
pages = {23-49},
year = {2017},
doi = {10.1146/annurev-fluid-010816-060322},
}

@article{malm_tubr_inflow,
title = {Spectral Element Simulation of Flow Around a
Surface-Mounted Square-Section Cylinder},
journal = {Proceedings of 20th Annual Conference of the CFD Society of Canada},
volume = {},
pages = {},
year = {2012},
issn = {},
doi = {},
url = {http://www.sinmec.ufsc.br/~dihlmann/MALISKA/proceedings_cfd_society_of_canada_conference_may_2012/papers/Malm_Schlatter_Henningson_Schrader_Mavriplis.pdf},
author = {Malm, J. and Schlatter, P. and Henningson, D. and
Schrader, L-U. and Mavriplis, C.}
}

@article{malm_tubr_inflow_journal,
title = {Direct numerical simulation of the flow around a wall-mounted square cylinder under various inflow conditions},
journal = {Journal of Turbulence},
volume = {},
pages = {},
year = {2015},
issn = {},
doi = {10.1080/14685248.2014.989232},
url = {https://doi.org/10.1080/14685248.2014.989232},
author = {R. Vinuesa and P. Schlatter and J. Malm and C. Mavriplis and D. Henningson}
}

@TechReport{nek5000_2021_tech_report,
   author = {Rezaeiravesh, S. and N. Jansson and
A. Peplinski and J. Vincent and P. Schlatter},
   title = {Nek5000: Theory, Implementation, Optimization},
   institution = {KTH Royal Institute of Technology},
   year = 2021,
url ={https://www.mech.kth.se/~pschlatt/Nek5000_2021.pdf}
}

@article{lyrintzis_inflow_review-2018,
author = {Dhamankar, N. S. and Blaisdell, G. A. and Lyrintzis, A. S.},
title = {Overview of Turbulent Inflow Boundary Conditions for Large-Eddy Simulations},
journal = {AIAA Journal},
volume = {56},
number = {4},
pages = {1317-1334},
year = {2018},
doi = {10.2514/1.J055528},

URL = { 
        https://doi.org/10.2514/1.J055528
}
}

@article{sillero_2013,
    author = {Sillero, J. A. and Jiménez, J. and Moser, R. D.},
    title = "{One-point statistics for turbulent wall-bounded flows at Reynolds numbers up to $\delta^+= 2000$}",
    journal = {Physics of Fluids},
    volume = {25},
    number = {10},
    pages = {105102},
    year = {2013},
    month = {10},
    issn = {1070-6631},
    doi = {10.1063/1.4823831},
    url = {https://doi.org/10.1063/1.4823831}
}

@article{kth_framework-2024,
title = {A comprehensive framework to enhance numerical simulations in the spectral-element code Nek5000},
journal = {Computer Physics Communications},
volume = {302},
pages = {109249},
year = {2024},
issn = {0010-4655},
doi = {https://doi.org/10.1016/j.cpc.2024.109249},
url = {https://www.sciencedirect.com/science/article/pii/S0010465524001723},
author = {D. Massaro and A. Peplinski and R. Stanly and S. Mirzareza and V. Lupi and T. Mukha and P. Schlatter}
}

@article{Inflow_inner_outer-Sandham-2003,
title = {Large-eddy simulation of transonic turbulent flow over a bump},
journal = {International Journal of Heat and Fluid Flow},
volume = {24},
number = {4},
pages = {584-595},
year = {2003},
note = {Selected Papers from the Fifth International Conference on Engineering Turbulence Modelling and Measurements},
issn = {0142-727X},
doi = {https://doi.org/10.1016/S0142-727X(03)00052-3},
url = {https://www.sciencedirect.com/science/article/pii/S0142727X03000523},
author = {N.D. Sandham and Y.F. Yao and A.A. Lawal}
}

@article{Perry_Chong_1982, title={On the mechanism of wall turbulence}, volume={119}, DOI={10.1017/S0022112082001311}, journal={Journal of Fluid Mechanics}, author={Perry, A. E. and Chong, M. S.}, year={1982}, pages={173–217}}

@article{Marusic_Perry_1995, title={A wall-wake model for the turbulence structure of boundary layers. Part 2. Further experimental support}, volume={298}, DOI={10.1017/S0022112095003363}, journal={Journal of Fluid Mechanics}, author={Marušić, I. and Perry, A. E.}, year={1995}, pages={389–407}}

@article{Perry_Marusic_1995, title={A wall-wake model for the turbulence structure of boundary layers. Part 1. Extension of the attached eddy hypothesis}, volume={298}, DOI={10.1017/S0022112095003351}, journal={Journal of Fluid Mechanics}, author={Perry, A. E. and Marušić, I.}, year={1995}, pages={361–388}}

@book{townsend1956,
  title={The structure of turbulent shear flow},
  author={Townsend, A.A.R.},
  year={1956},
  publisher={Cambridge university press}
}

@book{townsend1976,
  title={The structure of turbulent shear flow},
  author={Townsend, A.A.R.},
  year={1976},
  publisher={Cambridge university press}
}

@article{touber2009,
  title={Large-eddy simulation of low-frequency unsteadiness in a turbulent shock-induced separation bubble},
  author={Touber, E. and Sandham, N. D.},
  journal={Theoretical and computational fluid dynamics},
  volume={23},
  pages={79--107},
  year={2009},
  publisher={Springer}
}

@inbook{attached_eddy_turb_inflow-Marusic2006,
author = {Subbareddy, P. and Peterson, D. and Candler, G. and Marusic, I.},
title = {A Synthetic Inflow Generation Method Using the Attached Eddy Hypothesis},
booktitle = {24th AIAA Applied Aerodynamics Conference},
year={2006},
chapter = {},
pages = {},
doi = {10.2514/6.2006-3672},
URL = {https://arc.aiaa.org/doi/abs/10.2514/6.2006-3672}
}

@article{attached_eddy_review-marusic-2019,
author = {Marusic, I. and Monty, J. P.},
title = {Attached Eddy Model of Wall Turbulence},
journal = {Annual Review of Fluid Mechanics},
volume = {51},
number = {1},
pages = {49-74},
year = {2019},
doi = {10.1146/annurev-fluid-010518-040427},
URL = { 
        https://doi.org/10.1146/annurev-fluid-010518-040427
}
}

@article{dileep_chandran_2020,
  title = {Spectral-scaling-based extension to the attached eddy model of wall turbulence},
  author = {Chandran, D. and Monty, J. P. and Marusic, I.},
  journal = {Phys. Rev. Fluids},
  volume = {5},
  issue = {10},
  pages = {104606},
  numpages = {23},
  year = {2020},
  month = {Oct},
  publisher = {American Physical Society},
  doi = {10.1103/PhysRevFluids.5.104606},
  url = {https://link.aps.org/doi/10.1103/PhysRevFluids.5.104606}
}

@article{Sndberg_Sandham_2008, title={Direct numerical simulation of turbulent flow past a trailing edge and the associated noise generation}, volume={596}, DOI={10.1017/S0022112007009561}, journal={Journal of Fluid Mechanics}, author={Sandberg, R. D. and Sandham, N. D.}, year={2008}, pages={353–385}}

@article{Sandham_Sandberg-2009,
author = {N. D. Sandham and R. D. Sandberg},
title = {Direct numerical simulation of the early development of a turbulent mixing layer downstream of a splitter plate},
journal = {Journal of Turbulence},
volume = {10},
number = {},
pages = {N1},
year = {2009},
publisher = {Taylor \& Francis},
doi = {10.1080/14685240802698774},
URL = { https://doi.org/10.1080/14685240802698774}
}

@article{Stanly_Du-JFM-2024, title={Generating synthetic turbulence with vector autoregression of proper orthogonal decomposition time coefficients}, volume={1000}, DOI={10.1017/jfm.2024.1034}, journal={Journal of Fluid Mechanics}, author={Stanly, R. and Du, S. and Xavier, D. and Perez, A. and Mukha, T. and Markidis, S. and Rezaeiravesh, S. and Schlatter, P.}, year={2024}, pages={A83}}

@article{alfredsson_outer_peak-2011,
    author = {Alfredsson, P. H. and Segalini, A. and Örlü, R.},
    title = {A new scaling for the streamwise turbulence intensity in wall-bounded turbulent flows and what it tells us about the “outer” peak},
    journal = {Physics of Fluids},
    volume = {23},
    number = {4},
    pages = {041702},
    year = {2011},
    month = {04},
    issn = {1070-6631},
    doi = {10.1063/1.3581074},
    url = {https://doi.org/10.1063/1.3581074}
}

@article{universal_profile-Subrahmanyam_Cantwell_Alonso_2022, title={A universal velocity profile for turbulent wall flows including adverse pressure gradient boundary layers}, volume={933}, DOI={10.1017/jfm.2021.998}, journal={Journal of Fluid Mechanics}, author={Subrahmanyam, M. A. and Cantwell, B. J. and Alonso, J. J.}, year={2022}, pages={A16}}

@article{high_re_tbl_review-smits-2011,
   author = "Smits, A. J. and McKeon, B. J. and Marusic, I.",
   title = "High–Reynolds Number Wall Turbulence", 
   journal= "Annual Review of Fluid Mechanics",
   year = "2011",
   volume = "43",
   number = "Volume 43, 2011",
   pages = "353-375",
   doi = "https://doi.org/10.1146/annurev-fluid-122109-160753",
   url = "https://www.annualreviews.org/content/journals/10.1146/annurev-fluid-122109-160753",
   publisher = "Annual Reviews",
   issn = "1545-4479",
   type = "Journal Article",
  }

@article{synthetic_eddy_wall_info-pierre_sagaut-2009,
    author = {Pamiès, M. and Weiss, P-É. and Garnier, E. and Deck, S. and Sagaut, P.},
    title = {Generation of synthetic turbulent inflow data for large eddy simulation of spatially evolving wall-bounded flows},
    journal = {Physics of Fluids},
    volume = {21},
    number = {4},
    pages = {045103},
    year = {2009},
    month = {04},
    issn = {1070-6631},
    doi = {10.1063/1.3103881},
    url = {https://doi.org/10.1063/1.3103881}
}

@article{vonKarman1930,
  author = {Von Kármán, T.},
  title = {Mechanische Ähnlichkeit und Turbulenz},
  journal = {Nachrichten von der Gesellschaft der Wissenschaften zu Göttingen, Fachgruppe 1 (Mathematik)},
  year = {1930},
  pages = {58--76},
  language = {German}
}

@article{coles1956,
  author = {Coles, D.},
  title = {The law of the wake in the turbulent boundary layer},
  journal = {Journal of Fluid Mechanics},
  volume = {1},
  number = {2},
  pages = {191--226},
  year = {1956},
  doi = {10.1017/S0022112056000135}
}

@article{monkewitz2007,
  author = {Monkewitz, P. A. and Chauhan, K. A. and Nagib, H. M.},
  title = {Self-consistent high-Reynolds-number asymptotics for zero-pressure-gradient turbulent boundary layers},
  journal = {Physics of Fluids},
  volume = {19},
  number = {11},
  pages = {115101},
  year = {2007},
  doi = {10.1063/1.2779283}
}

@article{temporal_TBL-pasha-2024, title={Asymptotic scaling laws for periodic turbulent boundary layers and their numerical simulation up to $\textit{Re}_{\boldsymbol{\theta}}\text{ = 8300}$}, volume={1020}, DOI={10.1017/jfm.2025.10578}, journal={Journal of Fluid Mechanics}, author={Wynn, A. and Parvar, S. and O’Connor, J. and Frantz, R. A. S. and Laizet, S.}, year={2025}, pages={A6}}

@article{temporal_TBL-CnF-Biau2023,
title = {Self-similar temporal turbulent boundary layer flow},
journal = {Computers \& Fluids},
volume = {254},
pages = {105795},
year = {2023},
issn = {0045-7930},
doi = {https://doi.org/10.1016/j.compfluid.2023.105795},
url = {https://www.sciencedirect.com/science/article/pii/S0045793023000208},
author = {D. Biau}
}

@article{jansson2024neko,
  title={Neko: A modern, portable, and scalable framework for high-fidelity computational fluid dynamics},
  author={Jansson, N. and Karp, M. and Podobas, A. and Markidis, S. and Schlatter, P.},
  journal={Computers \& Fluids},
  volume={275},
  pages={106243},
  year={2024},
  publisher={Elsevier}
}

@BOOK{deville_fischer_mund_2002,
  title = {High-Order Methods for Incompressible Fluid Flow},
  publisher = {Cambridge University Press, Cambridge, U. K.},
  year = {2002},
  author = {Deville, M. O. and Fischer, P. F. and Mund, E. H.}
}

@article{Chauhan_2009,
doi = {10.1088/0169-5983/41/2/021404},
url = {https://dx.doi.org/10.1088/0169-5983/41/2/021404},
year = {2009},
month = {mar},
publisher = {},
volume = {41},
number = {2},
pages = {021404},
author = {K. A. Chauhan and P. A. Monkewitz and H. M. Nagib},
title = {Criteria for assessing experiments in zero pressure gradient boundary layers},
journal = {Fluid Dynamics Research}
}

@article{Simens_Jimenez2009,
title = {A high-resolution code for turbulent boundary layers},
journal = {Journal of Computational Physics},
volume = {228},
number = {11},
pages = {4218-4231},
year = {2009},
issn = {0021-9991},
doi = {https://doi.org/10.1016/j.jcp.2009.02.031},
url = {https://www.sciencedirect.com/science/article/pii/S0021999109001119},
author = {M. P. Simens and J. Jiménez and S. Hoyas and Y. Mizuno}
}

@article{Massaror_pipe-2024, title={Energy-based characterisation of large-scale coherent structures in turbulent pipe flows}, volume={996}, DOI={10.1017/jfm.2024.776}, journal={Journal of Fluid Mechanics}, author={Massaro, D. and Yao, J. and Rezaeiravesh, S. and Hussain, F. and Schlatter, P.}, year={2024}, pages={A45}}

@article{MONTY-pipe_channel_TBL-2009, title={A comparison of turbulent pipe, channel and boundary layer flows}, volume={632}, DOI={10.1017/S0022112009007423}, journal={Journal of Fluid Mechanics}, author={Monty, J. P. and Hutchins, N. and NG, H. C. H. and Marusic, I. and Chong, M. S.}, year={2009}, pages={431–442}}

@article{MONTY_pipe_channel-2007, title={Large-scale features in turbulent pipe and channel flows}, volume={589}, DOI={10.1017/S002211200700777X}, journal={Journal of Fluid Mechanics}, author={Monty, J. P. and Stewart, J. A. and Williams, R. C. and Chong, M. S.}, year={2007}, pages={147–156}}

@article{HUTCHINS_MARUSIC_2007, title={Evidence of very long meandering features in the logarithmic region of turbulent boundary layers}, volume={579}, DOI={10.1017/S0022112006003946}, journal={Journal of Fluid Mechanics}, author={Hutchins, N. and Marusic, I.}, year={2007}, pages={1–28}}

@article{scaling_mean_V_ZPG-2023, title={New formulations for the mean wall-normal velocity and Reynolds shear stress in a turbulent boundary layer under zero pressure gradient}, volume={969}, DOI={10.1017/jfm.2023.541}, journal={Journal of Fluid Mechanics}, author={Wei, Tie and Li, Zhaorui and Wang, Yanxing}, year={2023}, pages={A3}}

@book{trefethen2000spectral,
  title={Spectral Methods in MATLAB},
  author={Trefethen, L. N.},
  year={2000},
  publisher={Society for Industrial and Applied Mathematics},
  address={Philadelphia, PA}
}

@misc{fft_orthonormal_bounchaleun2019,
  title        = {An Elementary Introduction to Fast Fourier Transform Algorithms},
  author       = {A. Bounchaleun},
  year         = {2019},
  note         = {University of Chicago REU Paper}
}

@misc{perez2025_pysemtools,
      title={PySEMTools: A library for post-processing hexahedral spectral element data}, 
      author={A. Perez and S. Toosi and Olsen, T. F. and S. Markidis and P. Schlatter},
      year={2025},
      eprint={2504.12301},
      archivePrefix={arXiv},
      primaryClass={physics.comp-ph},
      url={https://arxiv.org/abs/2504.12301}, 
}

@article{jimenez1998largest,
  title={The largest scales of turbulent wall flows},
  author={Jim{\'e}nez, J.},
  journal={CTR Annual Research Briefs},
  volume={137},
  pages={54},
  year={1998},
  publisher={Stanford University}
}

\end{document}